\documentclass[twocolumn]{aastex7}

\makeatletter
\renewcommand{\frontmatter@title@above}{}
\makeatother

\usepackage{threeparttable}
\usepackage{graphicx}
\usepackage{xcolor}
\usepackage{xspace}
\usepackage{subfigure}
\usepackage{booktabs}
\usepackage{hyperref}
\usepackage{wrapfig}
\usepackage{sidecap}
\usepackage{float}
\usepackage{placeins}
\usepackage{capt-of}
\usepackage[version=4]{mhchem}

\hypersetup{
    hypertexnames=false,
    pdftitle={A Clearer View of HAT-P-1 b: JWST NIRSpec G395H Reveals Water, Carbon Dioxide, and Possibly Hydrogen Sulfide},
    pdfauthor={Reza Ashtari et al.}
}

\newcommand{\eureka}{\texttt{Eureka!}\xspace}
\newcommand{\firefly}{\texttt{FIREFLy}\xspace}
\newcommand{\Tswift}{\texttt{Tswift}\xspace}
\newcommand{\poseidon}{\texttt{POSEIDON}\xspace}
\newcommand{\sansar}{\texttt{SANSAR}\xspace}
\newcommand{\planetname}{HAT-P-1\,b}

\begin{document}

\onecolumngrid

\begin{center}
    {\textbf{
    A Clearer View of HAT-P-1\,b:\\
    JWST NIRSpec G395H Reveals Water, Carbon Dioxide,
    and Possibly Hydrogen Sulfide \\ \vspace{0.2cm}
    }}
\end{center}

\begin{center}
    Reza Ashtari$^{1}$,
    Stephen P. Schmidt$^{2,\ast}$,
    Guangwei Fu$^{2}$,
    Avinash Verma$^{3,4}$,
    David K. Sing$^{2,5}$,\\
    Kevin B. Stevenson$^{1}$,
    Jayesh Goyal$^{3,4}$,
    Katherine A. Bennett$^{5}$,
    Joshua D. Lothringer$^{6}$,
    Jacob Lustig-Yaeger$^{1}$,\\
    Sagnick Mukherjee$^{7,\dagger}$,
    Carlos Gascón$^{6}$,
    Natalie H. Allen$^{2}$,
    Patrick McCreery$^{2}$,
    Le-Chris Wang$^{8}$,\\
    Mei Ting Mak$^{9,10,\ddagger}$,
    Kristin S. Sotzen$^{1}$,
    Lakeisha M. Ramos Rosado$^{2}$,
    and N. J. Mayne$^{10}$
\end{center}

\begin{center}
\textit{
$^{1}$Johns Hopkins University Applied Physics Laboratory,
Laurel, MD, USA\\
$^{2}$William H. Miller III Department of Physics and Astronomy,
Johns Hopkins University, Baltimore, MD, USA\\
$^{3}$School of Earth \& Planetary Sciences,
National Institute of Science Education and Research (NISER),
Jatni, Odisha, India\\
$^{4}$Homi Bhabha National Institute, Mumbai, India\\
$^{5}$Morton K. Blaustein Department of Earth \& Planetary Sciences,
Johns Hopkins University, Baltimore, MD, USA\\
$^{6}$Space Telescope Science Institute, Baltimore, MD, USA\\
$^{7}$School of Earth and Space Exploration,
Arizona State University, Tempe, AZ, USA\\
$^{8}$Department of Astrophysical Sciences,
Princeton University, Princeton, NJ, USA\\
$^{9}$Atmospheric, Oceanic, and Planetary Physics,
University of Oxford, Oxford, UK\\
$^{10}$Department of Physics and Astronomy,
University of Exeter, Exeter, UK
}
\end{center}

\begin{center}
\textit{
$^{\ast}$NSF Graduate Research Fellow
\qquad
$^{\dagger}$51 Pegasi b Fellow
\qquad
$^{\ddagger}$Croucher Postdoctoral Fellow
}
\end{center}


\vspace{0.3cm}
\begin{center}
    \textit{Accepted to AJ; Revision Pending}
\end{center}

\vspace{0.4cm}

\begin{abstract}
\noindent
As part of JWST's Exoplanet Grand Tour Survey, we use panchromatic transmission spectroscopy to connect \planetname{}'s previously studied optical and near-infrared atmosphere to the longer-wavelength molecular bands accessible with JWST. We present JWST NIRSpec G395H transmission spectroscopy of the hot Jupiter \planetname{} over 2.7--5.3~$\mu$m, and combine the new spectrum with archival HST STIS and WFC3 observations for a 0.3--5.3~$\mu$m atmospheric analysis. We independently reduce the JWST data with the \eureka, \firefly, and \Tswift pipelines, finding mutually consistent transmission spectra across the G395H bandpass. Atmospheric retrievals yield strong evidence for \ce{H$_2$O} and \ce{CO$_2$} with Bayes factors of $\log_{10}\mathcal{B}_{\ce{H_2O}}=8.9$ and $\log_{10}\mathcal{B}_{\ce{CO_2}}=52.3$, while providing tentative evidence for \ce{H$_2$S} ($\log_{10}\mathcal{B}_{\ce{H_2S}}=1.4$). The joint \ce{H2O} and \ce{CO_2} constraints favor an atmosphere near chemical equilibrium, with $\log_{10} \text{M/H}=0.99^{+0.19}_{-0.14}$, corresponding to $\sim10\times$ Solar or  $\sim9\times$ relative to the near-solar metallicity host star, and a 3$\sigma$ upper limit of C/O $<0.52$. Because \ce{H2O} and \ce{CO2} provide a metallicity comparatively insensitive to vertical mixing in this temperature regime, their combined detection suggests the composition is dominated by bulk enrichment rather than strong disequilibrium transport.  We find no significant evidence for clouds; instead, the persistence of molecular structure across the spectrum argues against strong cloud muting. The tentative \ce{H2S} signal, if confirmed, would further suggest limited photochemical processing at the pressures probed. Together, the molecular inventory, enriched metallicity, and low C/O ratio point to an oxygen-rich atmosphere and establish \planetname{} as a benchmark for comparative studies of hot-Jupiter atmospheric composition.



\end{abstract}


\NewPageAfterKeywords
\section{Introduction} 
\label{sec:intro}

\planetname{} is a benchmark hot Jupiter ($T_{\mathrm{eq}}\sim1325$ K) whose system architecture and atmosphere have been studied extensively. The planet was identified by the HATNet survey as a low-density giant transiting one of the stars of a binary pair of G-stars with a projected separation of $\sim 1550$ AU \citep{Bakos2007}. The discovery observations established \planetname{} as a short-period ($P \simeq 4.465$ days), low-surface gravity planet with $M_p \simeq 0.53 M_J$ and $R_p \simeq 1.36 R_J$ \citep{Bakos2007}. 

Unless otherwise stated, the stellar and planetary properties adopted throughout this work are taken from the homogeneous analysis of the JWST Exoplanet Grand Tour targets by \citet{McCreery2026}.


Atmospheric observations of \planetname{} have accumulated a wealth of infrared and optical measurements across multiple telescopes. Spitzer Infrared Array Camera  (IRAC) eclipse photometry across the 3.6, 4.5, 5.8, and 8.0\,$\mu$m bands provided early insights into the temperature structure on the dayside and found the eclipse timing to be consistent with a near-circular orbit \citep{Todorov2010_Spitzer}. In transmission, Hubble Space Telescope (HST) Space Telescope Imaging Spectrograph (STIS) observations over $\sim0.3$--$1.0\,\mu$m found \ce{Na} absorption and strong optical absorption but found no significant \ce{K} feature, while HST Wide Field Camera 3 (WFC3) G141 observations ($\sim1.1$--$1.7\,\mu$m) measured a 1.4\,$\mu$m \ce{H$_2$O} band at 5$\sigma$ significance \citep{Nikolov14, Wakeford13}.

Ground-based optical spectroscopy and narrow-band photometry have added complementary constraints, including evidence for \ce{K} from GTC OSIRIS \citep{Wilson2015}, additional evidence from TNG DOLORES and Gemini GMOS \citep{Montalto2015, Todorov2010_Spitzer}, and detections of \ce{Na} and \ce{K} with P200 DBSP \citep{Chen2022}. Together, these studies make \planetname{} one of the more thoroughly-characterized hot Jupiters from optical through mid-infrared, and motivate a new analysis using James Webb Space Telescope (JWST) that can connect the HST-era molecular inventory to longer wavelength molecular tracers. 

Motivated by developing a consistent framework for comparing the atmospheres of giant exoplanets, the JWST Exoplanet Grand Tour Spectroscopic Survey (GO 5924) is designed as a first-generation JWST-HST program \citep{Sing2024_GT}. This legacy-quality set of transmission spectra spans the warm Saturn to hot Jupiter exoplanet regimes, assembled through complementary wavelength coverage intended to synthesize heterogeneous, cross-generational results. 

Here, we present JWST Near-infrared Spectrograph (NIRSpec) G395H transmission spectroscopy of \planetname{} obtained in Bright Object Time Series (BOTS) mode, which extends transmission constraints up to $\sim5\mu$m with time-series stability tailored to exoplanet transits. This wavelength range provides new insight into the relative abundances of water and carbon-bearing species that are difficult to disentangle or constrain with the sensitivity and shorter wavelength coverage provided by HST and Spitzer observations. Combining the new G395H spectrum with existing optical and NIR measurements, we refine the atmospheric composition of \planetname{} in the context of the Grand Tour survey.

We describe the JWST NIRSpec observing strategy and the construction of the G395H transmission spectrum in \autoref{sec:data}. We then present three independent reductions using \eureka, \firefly, and \Tswift in \autoref{sec:data:reduction}, followed by the white-light and spectroscopic light-curve analyses in \autoref{subsec:data:white_lc_fit} and \autoref{subsec:data:spec_lc_fit}. In \autoref{sec:modeling}, we interpret \planetname{}'s panchromatic transmission spectrum, combining HST STIS, HST WFC3, and JWST NIRSpec G395H data using atmospheric retrievals with \poseidon and \sansar, including free-chemistry, equilibrium-chemistry, and Bayesian model-comparison analyses. In \autoref{sec:discuss}, we discuss the implications of the retrieved molecular abundances and tentative sulfur-bearing opacity for the atmosphere of \planetname{} and for comparative studies of giant exoplanets. We summarize our conclusions in \autoref{sec:conclusion}.



\section{Data}
\label{sec:data}

\subsection{Observing Strategy}
\label{sec:data:observing_strategy}
\planetname ~was observed with NIRSpec BOTS, using the G395H grating, capturing the $\sim 2.7–5.3 ~\mu$m spectral range at $R\sim3000$. The observing setup employed the SUB2048 subarray. To mitigate saturation, the integration configuration consisted of 19 groups per integration (1336 total integrations), yielding a $\sim 6.7$ hr-long total exposure time for the transit observation. 

\subsection{Data Reduction}
\label{sec:data:reduction}
To assess the robustness and reproducibility of the G395H transmission spectrum of \planetname{}, we performed three fully independent data reductions using separate pipelines: \eureka \citep{Bell2022}, \firefly \citep{Rus22, Rus23}, and \Tswift \citep{fuWaterEscapingHelium2022}. Each reduction was carried out with its own choices regarding calibration, systematic-noise mitigation, background subtraction, spectral extraction, and light-curve fitting, thereby providing a check against reduction-dependent biases. All three reductions use the standard \texttt{jwst} ramp-fitting algorithm to convert detector ramps into count-rate products; the pipelines differ instead in the stage at which corrections such as $1/f$ subtraction are applied and in their subsequent background, extraction, and light-curve treatments.

For each reduction, white and spectroscopic light curves were fit independently to derive orbital and transit parameters. A comparison of these resulting parameters for all three pipelines is presented in \autoref{tab:orbital_parameters}. All parameters demonstrate good agreement across all fitted parameters, with differences well within the reported uncertainties.

The resulting transmission spectra from all three reductions are compared directly in \autoref{fig:comparison_spectra}. The spectra exhibit consistent transit depth and spectral structure across the full G395H wavelength range, with no evidence for discrepancies between reductions. This strong cross-pipeline consistency demonstrated the final G395H transmission spectrum is robust to data reduction choices and provides a reliable basis for atmospheric interpretation. 

\subsubsection{Eureka! Reduction}
\label{sec:data:reduction:eureka}

Our primary data reduction was performed using the \eureka pipeline, a modular JWST time-series spectroscopy framework designed for exoplanet transit and eclipse observations \citep{Bell2022}. Beginning with uncalibrated \texttt{uncal.fits} products, we processed the NRS1 and NRS2 detector data independently through Stages 1-4 of \eureka{}.  In Stage 1, we applied detector-level calibrations to the group-level integrations, including saturation flagging, trace curvature correction, dark-current mitigation, and ramp-fitting to produce count-rate images for each integration. In Stage 2, we performed flat-fielding and unit conversions. In Stage 3, we applied the wavelength solution, background subtractions to remove background fluxes outside the spectral trace, cleaned remaining hot pixels, traced the stellar spectrum independently for each detector, and extracted one-dimensional spectra using a box aperture centered on the spectral trace. In Stage 4, the extracted spectra were binned into white and spectroscopic channels and converted into time-series light curves with propagated uncertainties. We did not apply a separate $1/f$ correction in the \eureka reduction. Instead, the optimized column-by-column background treatment removes the dominant column-correlated component using pixels outside the trace. The agreement in both absolute transit depth and spectral shape with \firefly and \Tswift, which apply explicit $1/f$ corrections, indicates that any residual $1/f$ structure does not measurably bias the final 20-nm-binned spectrum. Flat-fielding is included in the \eureka Stage~2 processing and in the default \texttt{jwst} Stage~2 processing used by \firefly; the \Tswift implementation used here does not apply a separate flat-field correction before extraction. The close agreement among the three spectra suggests that this difference is negligible at the precision and binning of the present analysis.

For the \eureka reduction, we used the automated parameter-by-parameter optimization framework of \citet{Ashtari2025}. Key reduction parameters governing background estimation, outlier rejection, mask expansion, smoothing, and spectral extraction geometry were varied sequentially for NRS1 and NRS2. Here, ``smoothing'' refers to the \texttt{window\_len} parameter used when constructing the median spatial profile and identifying extraction outliers; it does not smooth the final transmission spectrum. The quality of each trial reduction was evaluated using the median absolute difference (MAD) of the resulting spectroscopic light curves. For each parameter, the value that minimized the MAD was adopted before proceeding to the next optimization step, yielding a sequentially optimized reduction designed to suppress point-to-point scatter without manually tuning the final transmission spectrum. The full optimization history for the \eureka reduction is shown in \autoref{fig:optimization_history}.

\subsubsection{\texttt{FIREFLy} Reduction}
\label{sec:data:reduction:firefly}

Our \texttt{FIREFLy} reduction \citep{Rus22,Rus23} of the JWST NIRSpec observations of HAT-P-1\,b is similar to the procedure used in \citet{Schmidt25}.
In Stage 1 we follow the default \texttt{jwst} pipeline steps.
As these observations also have fewer than 25 groups per integration, we skip the jump step. For short ramps, the group-to-group slope is less well constrained, which can cause the jump algorithm to flag an excessive number of false-positive cosmic rays; following \citet{Schmidt25}, we therefore identify temporal outliers at later stages instead.
We apply an integration-level $1/f$ subtraction in Stage 2 but otherwise follow the default \texttt{jwst} pipeline.
In Stage 3 we clean bad pixels by flagging ones with sharp variance spikes of over 100$\sigma$ using \texttt{lacosmic} \citep{lacosmic} and other known bad pixels in NIRSpec G395H as the first part of the stellar extraction. An aperture width of 4.76 pixels was used to extract the spectrum. This width was selected by minimizing the out-of-transit white-light residual scatter over a grid of trial apertures. The modified box extraction includes the fractional contribution of pixels intersected by the aperture boundaries, permitting non-integer aperture widths.
In the spectrophotometric light curve fitting stage of the \texttt{FIREFLy} pipeline, we begin by trimming the first 504 and last 10 pixel columns in NRS1 as well as the first 8 and last 18 pixel columns in NRS2.
We include both a linear and cubic trend in time as systematics in our \texttt{batman} \citep{kreidbergBatmanBAsicTransit2015} white light curve fit as preferred by the Bayesian Information Criterion (BIC). We compared polynomial baseline models containing combinations of terms through sixth order in centered time.
We report our best-fit white light curve parameters in Table \ref{tab:orbital_parameters} as determined by our Markov Chain Monte Carlo (MCMC) fit using \texttt{emcee} \citep{foreman-mackeyEmceeMCMCHammer2013}. The quadratic limb-darkening coefficients are fitted in the white-light analysis using the $u_{+}=u_1+u_2$ and $u_{-}=u_1-u_2$ parameterization. For the spectroscopic fits, we fix the orbital parameters to the respective white-light values for NRS1 and NRS2. We retain the white-light systematic model coefficients in each channel and fix the wavelength-dependent limb-darkening coefficients to the \texttt{MPS-ATLAS} set 2 model stellar atmosphere \citep{kostogryz2022stellar, kostogryz2023mps} values generated with the \texttt{ExoTiC-LD} Python package \citep{grant2024exoticLDJoss}.

\subsubsection{Tswift Reduction}
\label{sec:data:reduction:fu}

The \Tswift reduction follows a procedure similar to that described in \citep{fuWaterEscapingHelium2022}. The reduction begins with \texttt{uncal.fits} files, which are processed using the default \texttt{jwst} pipeline to produce \texttt{darkcurrentstep.fits} files. We then perform a group-level, column-by-column 1/f noise subtraction. Next, the ramp fitting step is applied to generate \texttt{rampfitstep.fits} files. Bad pixels, including hot pixels and cosmic rays, are subsequently identified and flagged for each frame. This is accomplished by comparing each frame to the median frame across the time series. The stellar spectra are then extracted for each frame using a 5-pixel aperture centered on the spectral trace for both NRS1 and NRS2. The white light curve is fitted using \texttt{batman} \citep{kreidbergBatmanBAsicTransit2015} in combination with the \texttt{emcee} fitter \citep{foreman-mackeyEmceeMCMCHammer2013}. The best-fit parameters are listed in Table~\ref{tab:orbital_parameters}. The parameters $a/R_\star$ and inclination are then fixed during the spectroscopic light curve fitting to derive the wavelength-dependent $R_p/R_\star$. The white light and spectroscopic light curve fitting systematic models include a multiplicative constant and a linear slope. For limb darkening, the $u_1$ and $u_2$ coefficients of the quadratic law are fixed based on the \texttt{stagger} stellar model grid \citep{magicStaggergridGrid3D2015}, adopting $T_{\mathrm{eff}} = 5980,\mathrm{K}$, $\log g = 4.5$, and $\mathrm{[M/H]} = 0$. The final transmission spectrum is shown in Figure~\ref{fig:comparison_spectra}.

\subsection{White Light Curve Fitting} 
\label{subsec:data:white_lc_fit}

\begin{figure*}[htb!]
    \centering
    \includegraphics[width=0.99\linewidth, trim={0cm 0cm 0cm 0cm},clip]{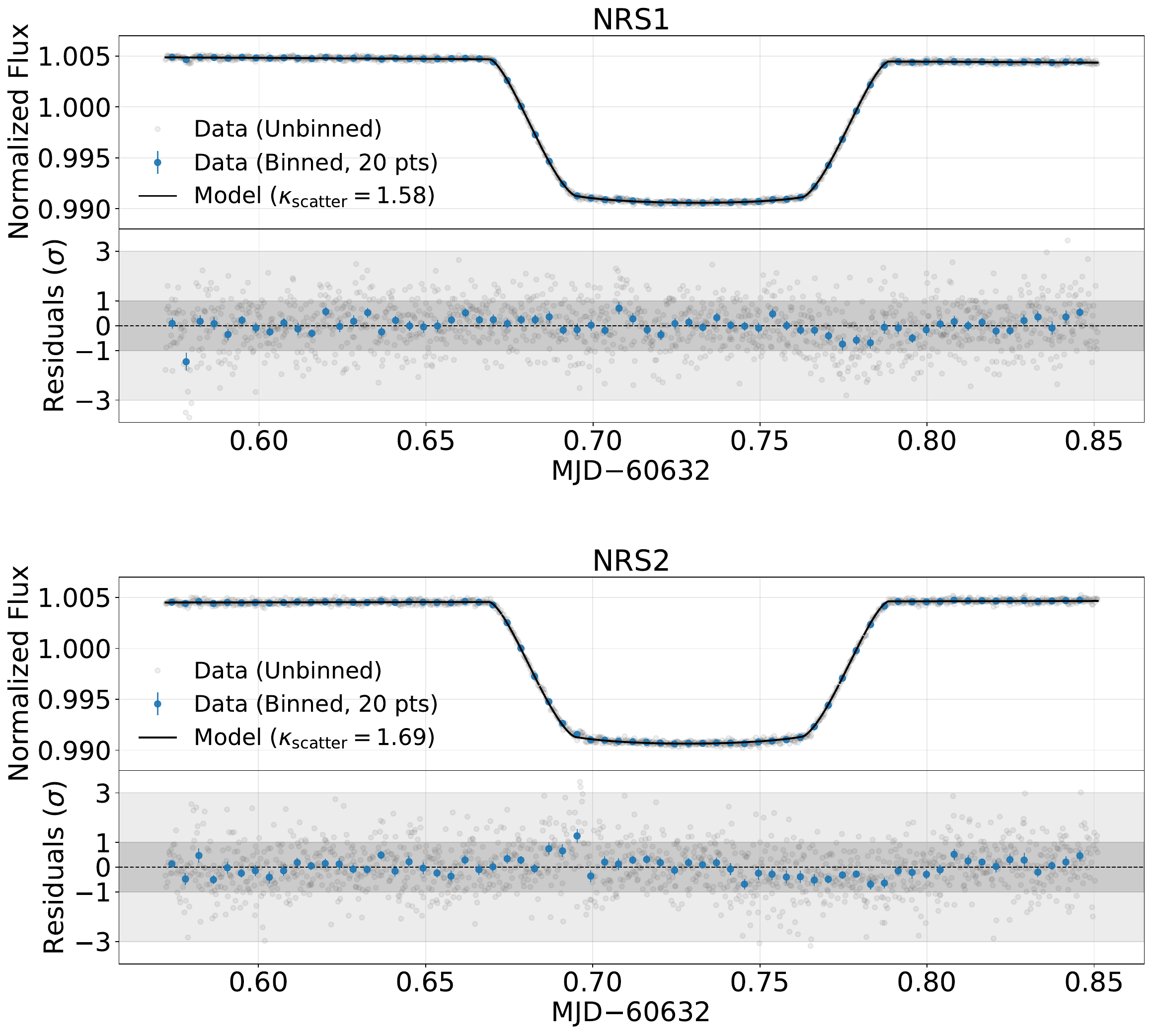} 
    \caption{G395H white-light transit fits for the two NIRSpec detectors, NRS1 (top) and NRS2 (bottom), derived from data reduced with the \eureka pipeline. In each panel, the upper subpanel shows the normalized flux time series, with unbinned data shown as gray points and time-binned data (20 points per bin) shown as blue points with uncertainties. Time is shown along the x-axis in MJD $-~60632$, where MJD = BJD $-~2400000.5$. The best-fitting transit model is overplotted as a solid black curve. The corresponding residuals, expressed in units of the photometric uncertainty ($\sigma$), are shown in the lower subpanels, with the shaded region indicating $\pm1\sigma$. The fitted scatter inflation factors ($\kappa_{\mathrm{scatter}}$) are reported in each panel legend as metrics for the transit model's fit to the \eureka light curve data, demonstrating statistically consistent fits for both detectors.
}
    \label{fig:wlc_fits}
\end{figure*}

\begin{figure*}[htb!]
    \centering
    \includegraphics[width=1.0\linewidth, trim={0.2cm 0cm 0.2cm 0cm},clip]{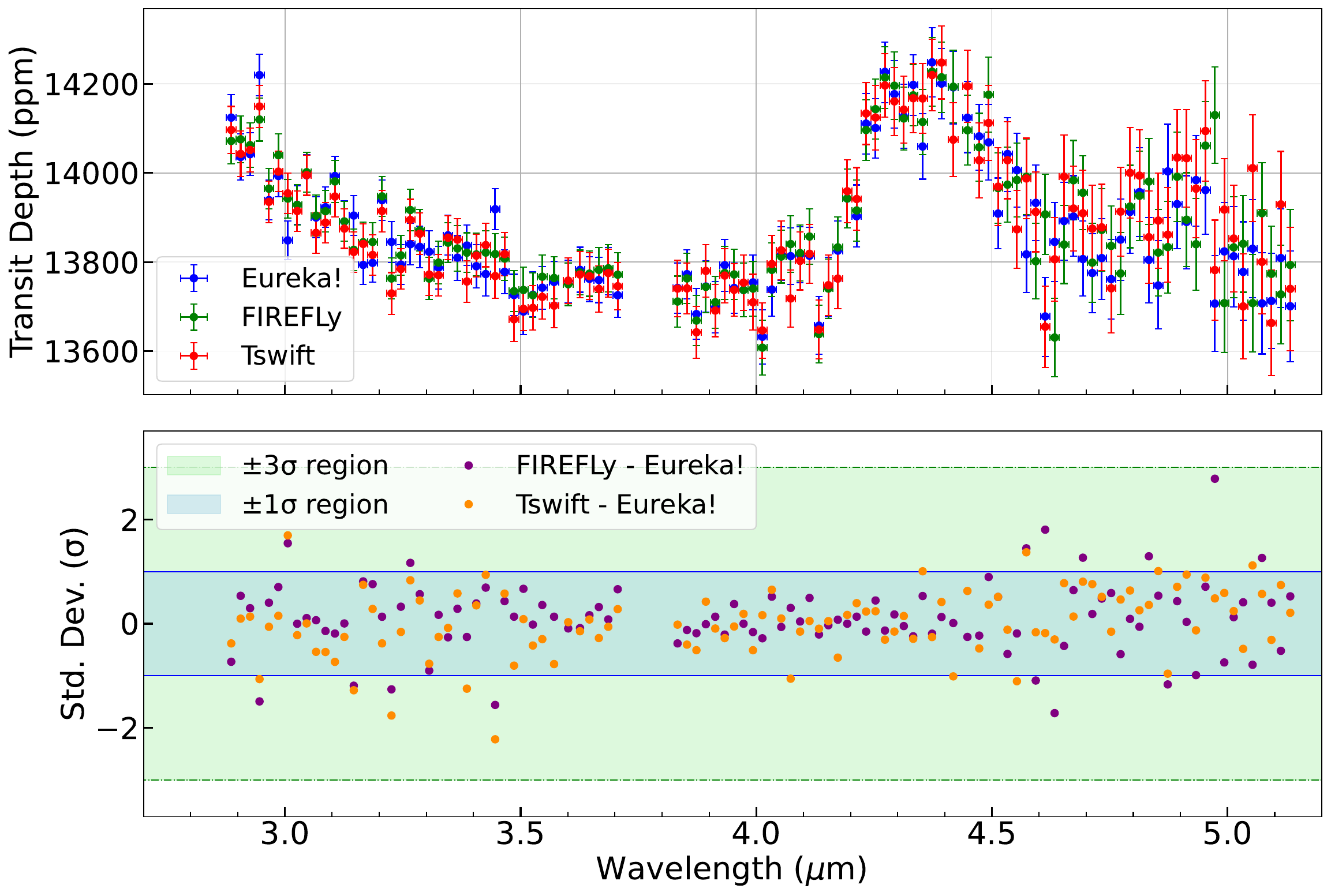} 
    \caption{
    Comparison of the NIRSpec G395H transmission spectra derived using the \eureka, \firefly, and \Tswift reduction pipelines. \textit{Top panel:} Transit depth (ppm) as a function of wavelength for each reduction, shown with 1$\sigma$ uncertainties. The three independently reduced spectra show excellent agreement in absolute transit depth and spectral structure across the full NRS1 and NRS2 wavelength range, with overlapping error bars for the vast majority of channels. \textit{Bottom panel:} Standardized residuals relative to \eureka, shown for \firefly$-\eureka$ (purple) and \Tswift$-\eureka$ (orange). The shaded regions indicate the $\pm1\sigma$ (blue) and $\pm3\sigma$ (green) intervals. Most spectral bins lie within $\pm1\sigma$, and nearly all fall within $\pm3\sigma$, with no evidence for systematic wavelength-dependent offsets, demonstrating strong cross-pipeline consistency and robust reproducibility of the G395H transmission spectrum.}
    \label{fig:comparison_spectra}
\end{figure*}

White-light transit fitting for the primary \eureka reduction was performed using a simultaneous joint fit to the NRS1 and NRS2 light curves. The white-light fitting procedures for the \firefly and \Tswift reductions are described in their respective reduction subsections; here, we focus on the \eureka fit used to define the fiducial orbital and transit parameters for the primary transmission spectrum. We fit the light curves with \texttt{emcee} using 400 walkers, 3000 steps, and 1000 burn-in steps, while allowing for detector-specific systematics and noise inflation. The best-fit white-light transit models and residuals are shown in \autoref{fig:wlc_fits}.

In the joint \eureka fit, the mid-transit time $T_0$, scaled semi-major axis $a/R_\star$, inclination $i$, and detector-specific planet-to-star radius ratios $R_p/R_\star$ were fit simultaneously. The orbital period was fixed to the value reported by \citet{Ment2018}, as a single transit observation does not provide an improved constraint on this parameter. All remaining instrumental and noise terms were allowed to vary independently for NRS1 and NRS2. The resulting best-fit transit parameters are compared with those from the \firefly and \Tswift reductions in \autoref{tab:orbital_parameters}.

\subsection{Spectroscopic Light Curve Fitting} 
\label{subsec:data:spec_lc_fit}

Spectroscopic light curves were generated for each wavelength channel and fit to extract the wavelength-dependent transit depth. For the primary \eureka reduction, the orbital parameters were fixed to the values obtained from the joint white-light fit described in \autoref{subsec:data:white_lc_fit}, while $R_p/R_\star$ was allowed to vary independently in each spectroscopic channel. 
Wavelength-dependent quadratic limb-darkening coefficients were computed with ExoTiC-LD using the MPS2 stellar atmosphere grid and applied as priors in the \eureka spectroscopic fits \citep{grant2024exoticLDJoss}. We adopted $T_{\mathrm{eff}}=5980$~K, $\log g=4.5$, and $\mathrm{[M/H]}=0.0$, corresponding to the nearest available grid point to the stellar parameters adopted from \citet{McCreery2026}. 

A diagnostic comparison between freely fitted spectroscopic limb-darkening coefficients against PHOENIX, STAGGER, and MPS2-predicted coefficients is shown in \autoref{fig:ld_comparison}; yielding a marginal preference for the MPS2 grid model as the best-fit to the freely fitted \eureka limb darkening coefficients.

The alternative \firefly and \Tswift spectroscopic fits follow the pipeline-specific procedures described in \autoref{sec:data:reduction:firefly} and \autoref{sec:data:reduction:fu}. We use the resulting independently reduced spectra as a cross-check on the primary \eureka transmission spectrum. The three G395H spectra are compared in \autoref{fig:comparison_spectra}, where they show consistent transit depths and spectral structure across both NRS1 and NRS2. This agreement demonstrates that the final G395H transmission spectrum is robust to differences in detector calibration, extraction choices, light-curve modeling, and limb-darkening treatment.


\section{Modeling} 
\label{sec:modeling}

\begin{figure*}[htb!]
    \centering
    \includegraphics[width=\linewidth]{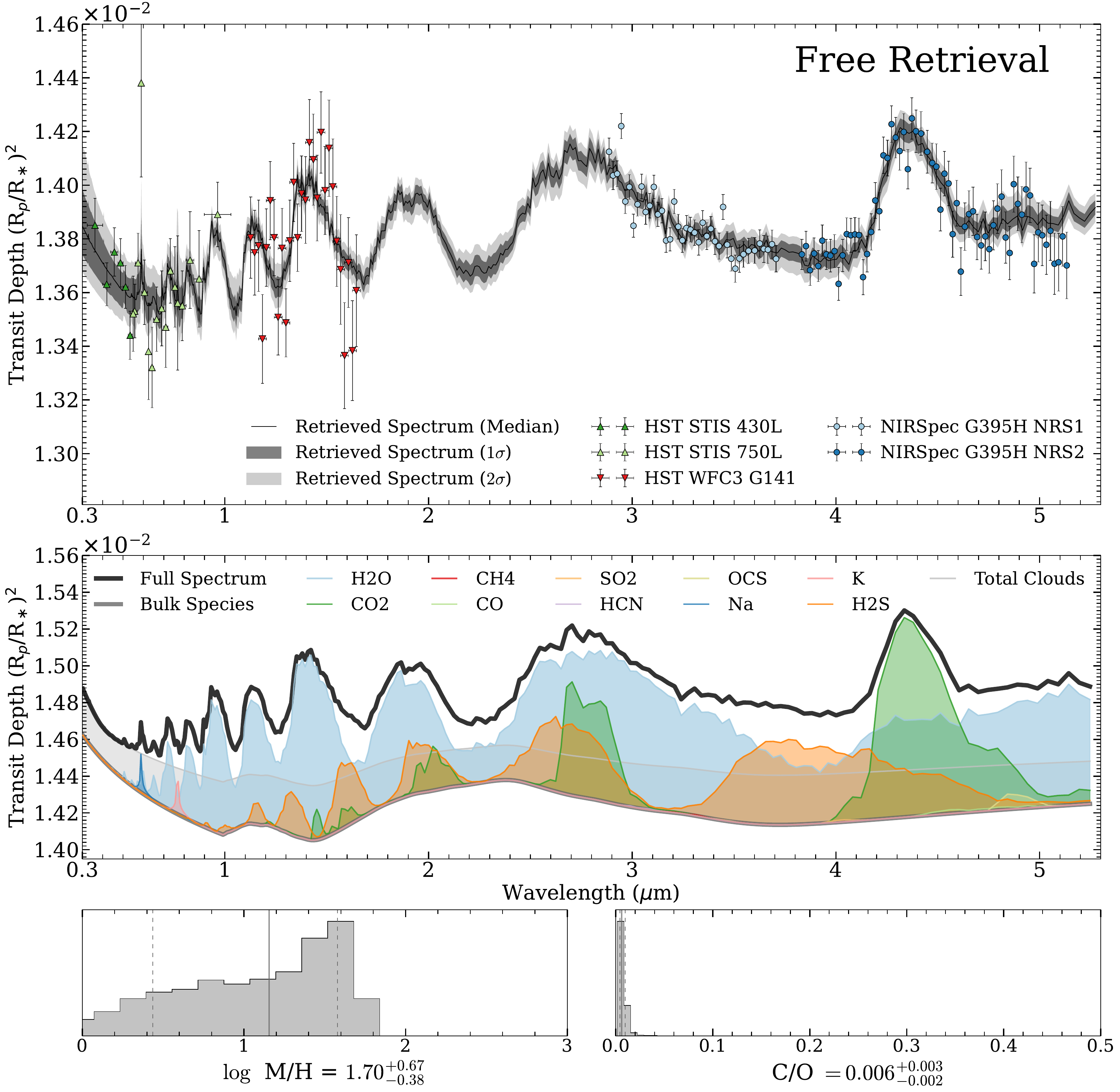}
    \caption{Summary of the results from our \poseidon panchromatic free retrieval. In the upper panel we show a comparison between the retrieved transmission spectrum and the data used in our analysis. 
    We plot as the black line the median retrieved spectrum and as (dark/light) gray polygons the (1/2)-$\sigma$ ranges of the retrieved spectrum.
    We overplot as colored points with black borders the HST (triangles) and JWST (circles) data points, differentiating between data from HST STIS (blue, upward-facing) and HST WFC3 (red, downward-facing) by the direction of the triangle as well as the color.
    We apply separate offsets between the JWST NIRSpec G395H \eureka reduction presented in this work and archival HST STIS and HST WFC3 data, with data from each instrument grouped together. 
    In the middle panel we show the spectral contribution corresponding to our median retrieved atmosphere, and we include beneath these panels the computed posteriors for the atmospheric metallicity and C/O ratio. For each free-retrieval posterior sample, C/O is calculated by stoichiometrically summing the carbon and oxygen atoms contained in the retrieved C- and O-bearing species. 
    The metallicity M/H is calculated as the ratio between the weighted average of the atmosphere's stoichiometric C/H, O/H, N/H, P/H, and S/H ratios and their respective \citet{Asplund2021} Solar values. 
    These derived quantities are informative summaries rather than independently retrieved parameters.
    Owing to the high H$_2$O VMR several orders of magnitude higher than any other carbon- or oxygen-containing molecule, the computed metallicity is quite high while the C/O ratio is very low.
    Due to the lack of a CO or CH$_4$ detection, these values have little physical meaning.
    We discuss the molecular detection significances and nested model comparisons related to this free retrieval in \autoref{sec:modeling:retrievals} and summarize them in \autoref{tab:stats}.
    }    \label{fig:panchromatic_free_retrieval}
\end{figure*}

\begin{figure*}[htb!]
    \centering
    \includegraphics[width=\linewidth]{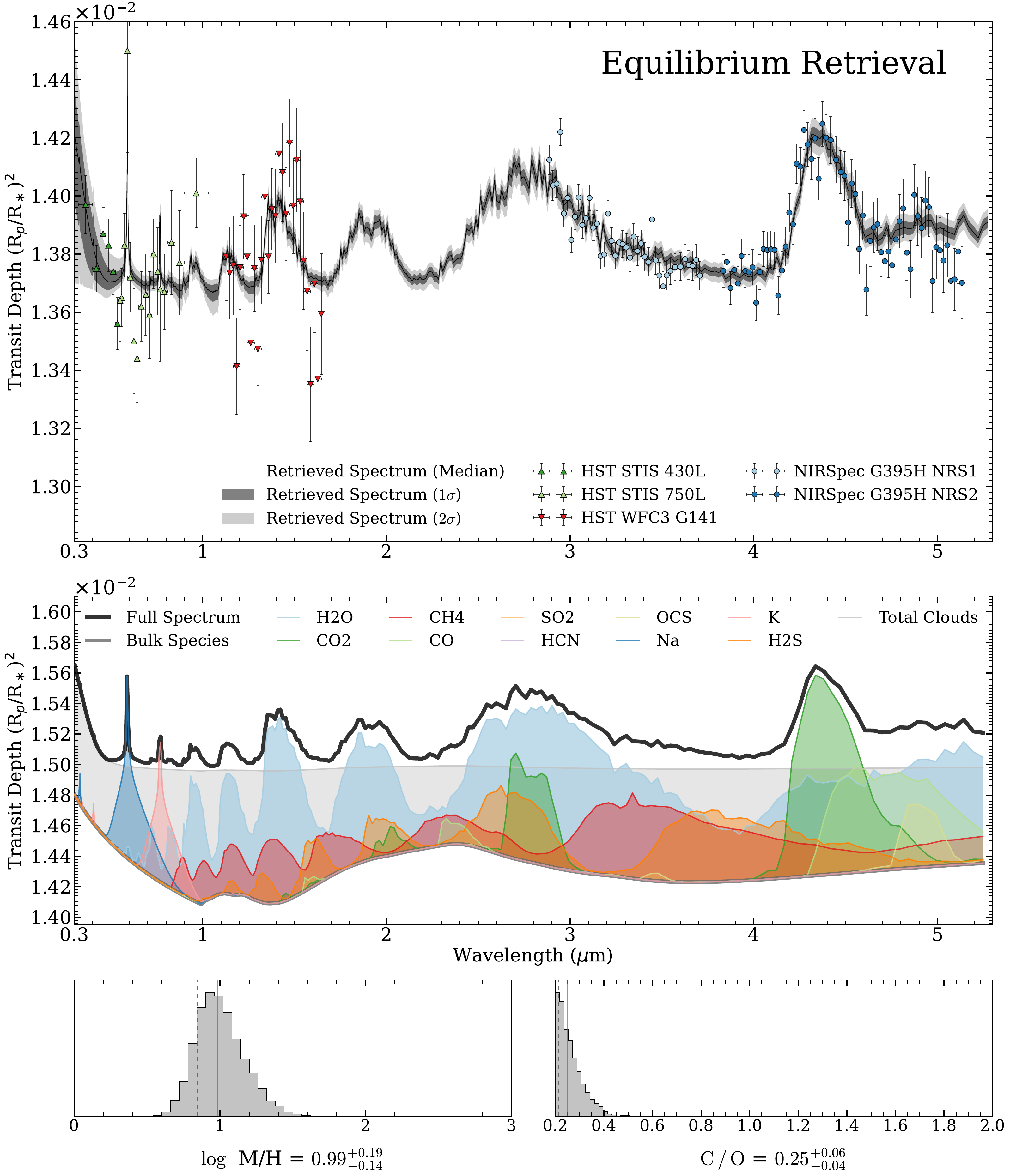}
    \caption{Summary of the results from our \poseidon panchromatic equilibrium chemistry retrieval. The format of the upper two panels of this figure is identical to that of Figure \ref{fig:panchromatic_free_retrieval}, but the metallicity and C/O ratio are derived from the equilibrium chemistry grid rather than computed from molecule volume mixing ratios. We find \planetname{}'s atmosphere to have a $\log_{10}\mathrm{[M/H]}=0.99^{+0.19}_{-0.14}$, corresponding to approximately $10\times$ Solar or $9\times$ the near-Solar host-star metallicity, and a factor of at least 4 more oxygen than carbon.
    There is a notable discrepancy in the cloud-coverage fraction between the free and equilibrium \poseidon retrievals; JWST NIRISS SOSS observations of \planetname{} will help alleviate this tension in the future.}
    \label{fig:panchromatic_eq_retrieval}
\end{figure*}

\subsection{Atmosphere Retrievals}
\label{sec:modeling:retrievals}

We perform a retrieval analysis on the transmission spectrum of \planetname{} using two retrieval codes: the open-source \poseidon code  \citep{MacDonald2023} and the new \texttt{SANSAR} retrieval code \citep{Verma_2025}. In addition to the JWST NIRSpec G395H observations presented in this article, we also incorporate archival transmission spectra taken by HST's WFC3 \citep[presented in][]{Wakeford13} and STIS \citep[presented in][]{Nikolov14} instruments in our retrievals to obtain a comprehensive view of \planetname{}'s atmosphere.

\subsubsection{\poseidon}
\label{sec:poseidon_model}

We use the \poseidon Python package \citep{MacDonald2017,MacDonald2023} to perform both free chemistry and equilibrium chemistry atmosphere retrievals on our panchromatic 0.3--5.3\,$\mu\text{m}$ transmission spectrum of \planetname{}.
We choose to perform an equilibrium chemistry retrieval in addition to a free retrieval as the major atmospheric species (like H$_2$O, CO, and CO$_2$) are expected to follow equilibrium chemistry for planets with equilibrium temperatures T$_{\text{eq}} \gtrsim 1100$ K \citep{Mukherjee2025}.
Given that planets with equilibrium temperatures only a few hundred Kelvin cooler than \planetname{} have been found to be experiencing disequilibrium chemistry \citep[e.g., WASP-39\,b;][]{Rus23, Tsai2023, Powell24}, the comparison between equilibrium and free chemistry will serve as a gauge of the degree to which \planetname{}'s atmosphere departs from equilibrium.
Our atmosphere model assumes background gas consisting of H$_2$ and He at Solar He/H$_2$ = 0.17 \citep{asplund09}.
We construct atmospheres with 150 layers uniformly spaced in log pressure, ranging from 10$^{-7}$--100 bar, under hydrostatic equilibrium.
We solve hydrostatic equilibrium with the boundary condition that the atmosphere reaches a pressure of 10$^{-3}$ bar at the reference radius.
We freely retrieve the $T(P)$ profile following the prescription of \citet{Madhusudhan2009} and fix the planet's surface gravity to 6.66 m s$^{-1}$ as derived from the adopted mass and radius in the homogeneous Grand Tour analysis of \citet{McCreery2026}.
Both our equilibrium chemistry and free chemistry atmosphere models consider the following trace species: \ce{H$_2$O}, \ce{CO$_2$}, CH$_4$, CO, SO$_2$, HCN, OCS, \ce{H$_2$S}, Na, and K.
For our equilibrium chemistry retrievals, we retrieve the atmospheric metallicity and C/O ratio, whereas for our free chemistry retrievals we retrieve the log volume mixing ratios (VMRs) for each of the trace species in our model.
For clouds and aerosols, we use the inhomogeneous cloud and haze parameterization of \citet{MacDonald2017} wherein aerosols are parameterized via a Rayleigh enhancement factor $\log a$ for the H$_2$-Rayleigh scattering cross-section, a scattering slope $\gamma$, a cloud-top pressure $\log P_{\text{cloud}}$, and a cloud-coverage fraction $\phi$.
In total, our equilibrium chemistry model has 13 free parameters: the reference radius, 6 $T(P)$ profile parameters, 4 cloud/haze parameters, and the atmospheric metallicity and C/O ratio as equilibrium chemistry parameters.
Our free chemistry retrievals replace the two equilibrium chemistry parameters with 10 individual trace species VMRs, increasing the number of parameters for these retrievals to 21.

We calculate model spectra in our \poseidon retrievals by solving the equation of radiative transfer in a cylindrical coordinate system for 100 incident stellar rays that are attenuated according to the atmospheric opacity along line-of-sight.
We pre-compute our opacities at $R=20,000$ across a grid of temperatures and pressures using the \texttt{Cthulhu} Python package \citep{Agrawal2024,Cthulu}.
We derive our opacities using the following spectroscopic line lists or measured cross sections: \ce{H$_2$O} \citep[POKAZATEL;][]{polyansky18},  \ce{CO$_2$} \citep[UCL-4000;][]{yurchenko20}, CH$_4$ \citep[MM;][]{Yurchenko2024}, CO \citep[Li2015;][]{li15}, SO$_2$ \citep[ExoAmes;][]{Underwood2016}, HCN \citep[Harris;][]{barber14},  OCS \citep[OYT8;][]{Owens2024}, \ce{H$_2$S} \citep[AYT2;][]{azzam16exomol}, Na and K \citep[VALD3;][]{Ryabchikova15}.
We also include continuum collision-induced absorption from H$_2$-H$_2$ and H$_2$-He pairs \cite{Karman2019} and Rayleigh scattering for all gases \cite{MacDonald2022}. 
We allow for two free offsets: one between NIRSpec G395H and HST STIS, and another between NIRSpec G395H and HST WFC3/G141. We do not include an additional offset between NRS1 and NRS2. Their independently fitted white-light radius ratios agree within $1\sigma$ (\autoref{tab:orbital_parameters}), and the transmission spectrum shows no coherent discontinuity across the detector gap (\autoref{fig:comparison_spectra}); an unconstrained detector offset would therefore add a largely degenerate parameter without evidence that it is required.
We calculate model transmission spectra using opacity sampling on a wavelength grid from 0.3--5.3$\mu$m, and we use nested sampling via the \texttt{PyMultiNest} \citep{Feroz2009,Buchner2014} Python package with 1000 live points to explore the parameter space for each of our retrievals.

To assess the statistical significance of the detections of molecules in the atmosphere of \planetname{}, we also perform a series of nested free retrievals.
In these retrievals we exclude one molecule of interest while keeping the rest of the configuration identical.
We perform these nested retrievals for \ce{H$_2$O}, \ce{CO$_2$}, and \ce{H$_2$S} individually.
We also perform a free retrieval without a cloud and haze treatment to assess the significance of the presence of clouds.

\subsubsection{\sansar}

 We use the transmission-spectrum module of \sansar \citep{Verma_2025} to perform an independent set of free-chemistry retrievals on the same panchromatic 0.3--5.2~$\mu$m HST/STIS, HST/WFC3, and JWST/NIRSpec data. The atmosphere is divided into 100 layers spaced uniformly in log pressure from 100~bar at the bottom to $10^{-8}$~bar at the top. We assume an H$_2$-dominated atmosphere and initialize the unnormalized H$_2$/He fractions to 0.84/0.15, corresponding to He/H$_2=0.178$. We include the same trace species used in the \poseidon analysis (\autoref{sec:poseidon_model}). Line-by-line cross sections are computed over pressure and temperature grids using the line lists and partition functions summarized in Table~8 of \citet{Verma_2025}, with broadening parameters from the supplementary material of \citet{Goyal2020}. We generated the OCS cross sections for this work with HELIOS-K using the HITRAN line list \citep{GORDON2022107949}. These cross sections were converted into correlated-$k$ tables at $R\sim1000$ for use in the transmission calculations.

 We fix the reference pressure to $10^{-2}$~bar and retrieve the planetary radius at that pressure, $R_{p,\mathrm{ref}}$. The free-chemistry retrieval adopts the six-parameter $P$--$T$ profile of \citet{Madhusudhan2009} and the four-parameter inhomogeneous cloud-and-haze model of \citet{MacDonald2017}. In this model, $\alpha_{\mathrm{haze}}$ sets the Rayleigh-scattering enhancement, $\gamma$ sets the scattering slope, $P_{\mathrm{top}}$ is the top pressure of a wavelength-independent gray cloud deck, and $\phi$ is the cloudy terminator fraction. We also retrieve two relative offsets: one between HST/STIS and NIRSpec/G395H and one between HST/WFC3 and NIRSpec/G395H. The \sansar model therefore has 23 free parameters. We sample the posterior and Bayesian evidence with \texttt{PyMultiNest} \citep{Feroz2009}, using 1000 live points and an evidence tolerance of 0.5.

\subsubsection{Retrieval Results}

Our free \poseidon retrievals show strong evidence for the presence of \ce{H$_2$O} and \ce{CO$_2$} in the atmosphere of \planetname{}, with Bayes factors $\log_{10}(\mathcal{B})_{\ce{H2O}} = 8.9$ and $\log_{10}(\mathcal{B})_{\ce{CO2}} = 52.3$. Both correspond with robust detections according to the Jeffries' Scale and ``Decisive'' evidence according to \citet{Thorngren26}. 
The \ce{CO$_2$} feature corresponds to a VMR $\log \text{CO}_2=-3.98^{+0.53}_{-0.66}$ and is constrained by the feature present at 4.3 microns \citep[as it has been for several other hot Jupiters, e.g.,][]{JWSTCO2, Fu24, Ahrer25, Kirk25, Meech25}.
Notably, this \ce{CO$_2$} VMR is among the highest in the literature for a hot Jupiter, probably due to the lack of detected CO (which is expected to be over an order of magnitude more abundant under the assumptions of equilibrium chemistry). 
The \ce{H$_2$O} feature, with a VMR $\log \text{H}_2\text{O}=-1.70^{+0.43}_{-0.72}$, is constrained by a wide feature in NRS1. 
Meanwhile, we find moderate-to-strong evidence for \ce{H$_2$S} given its Bayes factor of $\log_{10}(\mathcal{B})_{\ce{H2S}} = 1.4$.
Its VMR, $\log \text{H}_2\text{S}=-3.30^{+0.41}_{-0.65}$, also places it among the highest in the literature; however, it is constrained primarily by data points near the gap between NIRSpec G395H's detectors and shallower transit depths in the longest-wavelength HST WFC3/IR G141 data.
As an instrumental systematic may be biasing the G141 data redward of 1.5 $\mu$m as evidenced by lower values relative to higher-precision NIRISS SOSS data (e.g., Schmidt et al., in preparation), it is possible that our inference of \ce{H$_2$S} is less significant as a result.
 The cloud model is not significantly favored. Its small evidence improvement is distributed across the broadband continuum rather than being driven by a distinct cloud feature, the cloud posteriors are broad, and BPICs instead favors the cloud-free model (\autoref{tab:stats}). Together with the absence of a measurable morning--evening radius difference, we therefore do not interpret the 2D-cloud solution as compelling evidence for physically asymmetric cloud coverage.
Future NIRISS SOSS observations of \planetname{} will be able to distinguish between these possibilities more decisively.

Turning to our equilibrium chemistry model, our \poseidon equilibrium chemistry retrieval prefers an atmospheric metallicity of  $\log_{10}\mathrm{[M/H]}=0.99^{+0.19}_{-0.14}$, corresponding to approximately $10\times$ Solar or $9\times$ the near-Solar host-star metallicity adopted from \citet{McCreery2026}, for \planetname{}.
We infer a low C/O ratio, at the edge of the prior range and with a three-sigma upper limit $\text{C/O}<0.52$.
This is expected given the dominance of \ce{H$_2$O} and \ce{CO$_2$} in \planetname{}'s atmosphere.
While the metallicity is well-constrained thanks to the combination of strong H$_2$O and CO$_2$ features that are not strongly muted by aerosols, the C/O ratio is not precisely measured because we lack tight constraints on CH$_4$ and, more importantly, CO, which is expected to be the main carrier of carbon.
Given that CH$_4$ is expected to be out of equilibrium at these equilibrium temperatures \citep[e.g.,][]{Mukherjee2025}, it is not surprising that we do not detect it. 
However, the significant CO$_2$ feature does suggest a solar-to-sub-solar C/O ratio as implied by the chemical equilibrium retrievals. We did not perform a separate G395H-only retrieval as the spectral contributions show the principal molecular tracers are well-measured via panchromatic analysis: the 4.3~$\mu$m CO$_2$ constraint is supplied by G395H, the broad NRS1 structure provides substantial H$_2$O opacity, while HST data also help measure the 1.4~$\mu$m H$_2$O band. 

The independent \sansar free-chemistry retrieval provides a complementary check on the atmospheric interpretation inferred from \poseidon. The retrieved abundances and parameter constraints from \sansar are shown in \autoref{fig:sansar_panchromatic_retrieval} and \autoref{tab:retrieval_configurations}. Using the same panchromatic HST+JWST transmission spectrum and comparable free parameters and species, the constraints on \ce{H$_2$O} and \ce{CO$_2$} abundances in \sansar are comparable to that of \poseidon, noted from the agreement between uncertainties on the constrained abundances. We also retrieve \ce{H$_2$S} abundance as $-3.15_{-0.27}^{+0.24}$, which is very similar to \poseidon but with higher Bayes factor compared to the reference model, as shown in Table \ref{tab:stats}.  The retrieved cloud-top pressure is $\log_{10}(P_{\mathrm{cloud}}/\mathrm{bar})=-0.61^{+1.69}_{-2.64}$, with $\phi=0.29^{+0.39}_{-0.20}$. The broad cloud posteriors and lack of strong correlations with the molecular abundances are consistent with the absence of significant evidence for clouds. 



\begin{figure*}[t!] 
    \centering
    \includegraphics[width=0.99\linewidth, trim={0cm 0cm 0cm 0cm}, clip]{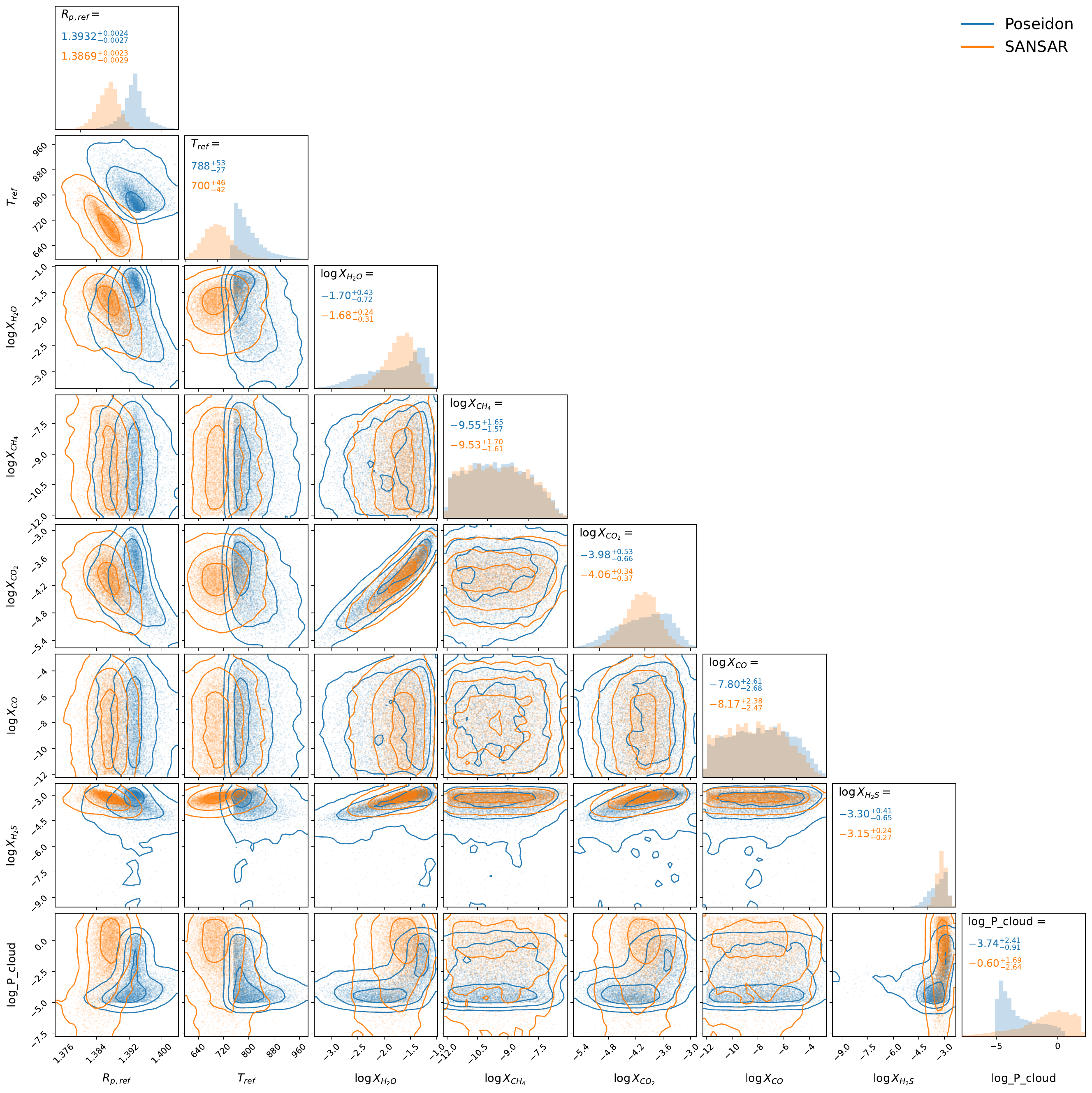}
    \caption{
    Overlaid posterior distributions from the primary \poseidon retrieval and the independent \sansar retrieval for a subset of shared free-chemistry parameters. One- and two-dimensional marginalized posteriors are shown for the reference planetary radius $R_{p,\mathrm{ref}}$, reference atmospheric temperature $T_{\mathrm{ref}}$, cloud top pressure $P_{\mathrm{cloud}}$, and volume mixing ratios of \ce{H$_2$O}, \ce{CH4}, \ce{CO$_2$}, \ce{CO} and \ce{H$_2$S}. Blue contours and histograms correspond to \poseidon, while orange contours and histograms correspond to \sansar. Values shown along the diagonal give the posterior medians and 1$-\sigma$ values. Both retrieval frameworks recover consistent \ce{H$_2$O}, \ce{CO$_2$}, and \ce{H$_2$S} abundances, while showing modest offsets in $R_{p,\mathrm{ref}}$ and $T_{\mathrm{ref}}$, reflecting differences in the forward-model implementations, prior choices, and retrieval assumptions. The cloud-top pressure posteriors are broad and not tightly constrained in either retrieval, with \poseidon favoring lower pressures than \sansar but both frameworks showing substantial uncertainty, consistent with the weak evidence for clouds.
    }
    \label{fig:overlay_corner}
\end{figure*}

\begin{deluxetable}{lcc}[b]
\tablewidth{0pt}
\tabletypesize{\footnotesize}
\tablecaption{
Posterior abundance constraints from the free-chemistry \poseidon and \sansar retrievals. 
For each species, we report the median and 1 - $\sigma$ values of the retrieved volume mixing ratio, expressed as $\log_{10} X_i$. 
\label{tab:vmr_posteriors}
}
\tablehead{
\colhead{Molecule} & 
\colhead{\poseidon $\log_{10} X_i$} & 
\colhead{\sansar $\log_{10} X_i$}
}
\startdata
\ce{H2O}  & $-1.70^{+0.43}_{-0.71}$ & $-1.68^{+0.24}_{-0.30}$ \\
\ce{CO2}  & $-3.99^{+0.53}_{-0.66}$ & $-4.07^{+0.34}_{-0.36}$ \\
\ce{CH4}  & $-9.55^{+1.65}_{-1.59}$ & $-9.56^{+1.73}_{-1.58}$ \\
\ce{CO}   & $-7.85^{+2.63}_{-2.61}$ & $-8.16^{+2.34}_{-2.44}$ \\
\ce{SO2}  & $-8.47^{+1.91}_{-2.22}$ & $-8.87^{+2.04}_{-2.01}$ \\
\ce{HCN}  & $-9.35^{+1.82}_{-1.77}$ & $-8.96^{+1.98}_{-1.98}$ \\
\ce{OCS}  & $-9.38^{+1.51}_{-1.73}$ & $-7.63^{+1.00}_{-2.25}$ \\
Na        & $-8.52^{+1.37}_{-1.92}$ & $-8.73^{+1.32}_{-1.94}$ \\
K         & $-9.74^{+1.65}_{-1.50}$ & $-9.63^{+1.60}_{-1.54}$ \\
\ce{H2S}  & $-3.30^{+0.41}_{-0.65}$ & $-3.15^{+0.24}_{-0.26}$ \\
\enddata
\end{deluxetable}

\begin{deluxetable*}{lcccccccc} 
    \renewcommand{\arraystretch}{1.5}
    \tabletypesize{\footnotesize}
    \tablecolumns{9} 
    \tablecaption{
    Retrieval statistics for the panchromatic HST$+$JWST transmission-spectrum retrievals.
    For each nested model, we report the difference in Bayesian evidence relative to the corresponding full model, $\Delta \ln(Z)$, the Bayes factor $\log_{10}(\mathcal{B})$, BPICS, reduced $\chi^2$, total $\chi^2$, degrees of freedom, and qualitative interpretation.
    Positive $\Delta \ln(Z)$ values indicate that the nested model is disfavored relative to the corresponding full model.
    \label{tab:stats}}
    \tablehead{
    Retrieval Type & Model & $\Delta \ln(Z)$ & $\log_{10}(\mathcal{B})$ & BPICs & $\chi^2_{\text{red}}$ & $\chi^2$ & DOF & Significance
    }
    \startdata
    \hline
    \multicolumn{9}{c}{\poseidon}\\
    \hline
    Free chemistry & All species & Reference & Reference & -2426.2 & 1.30 & 174.15 & 134  & ---\\
    Free chemistry & No \ce{H2O} & 20.52 & 8.91 & -2397.1 & 1.52 & 205.82 & 135 & Strong Detection\\
    Free chemistry & No \ce{CO2} & 120.43 & 52.30 & -2193.7 & 2.94 & 397.47 & 135 & Strong Detection\\
    Free chemistry & No \ce{H2S} & 3.20 & 1.39 & -2427.8 & 1.33 & 179.70 & 135 & Tentative Evidence\\
    Free chemistry & No Clouds & 1.79 & 0.78 & -2431.3 & 1.29 & 177.56 & 138 & No Evidence\\
    Equilibrium chemistry & All species & Reference & Reference & -2423.3 & 1.36 & 193.31 & 142 & ---\\
    \hline
    \multicolumn{9}{c}{\sansar}\\
    \hline
    Free chemistry & All species & Reference & Reference & \nodata & 1.137 & 152.41 & 134 & ---\\
    Free chemistry & No \ce{H2O} & 26.59 & 11.55 & \nodata & 1.471 & 198.61 & 135 & Strong Detection\\
    Free chemistry & No \ce{CO2} & 127.64 & 55.43 & \nodata & 3.064 & 413.66 & 135 & Strong Detection\\
    Free chemistry & No \ce{H2S} & 6.66 & 2.89 & \nodata & 1.281 & 172.94 & 135 & Weak Evidence\\
    \enddata
    \tablecomments{
    $\Delta \ln(Z)$ is computed relative to the corresponding full model within each retrieval framework, such that $\Delta \ln(Z)=\ln(Z)_{\rm full}-\ln(Z)_{\rm nested}$. 
    The listed $\log_{10}(\mathcal{B})$ values are equivalent to $\Delta \ln(Z)/\ln(10)$. 
    BPICS is included as an information-criterion check on prior sensitivity following \citet{Thorngren26}; lower BPICS values indicate better predictive performance.
    Absolute evidences and BPICS values should not be compared directly between \poseidon and \sansar because the two frameworks differ in forward-model assumptions, opacity treatment, priors, and implementation details.
    }
\end{deluxetable*}

\begin{figure*}
     \centering
    \includegraphics[width=\linewidth]{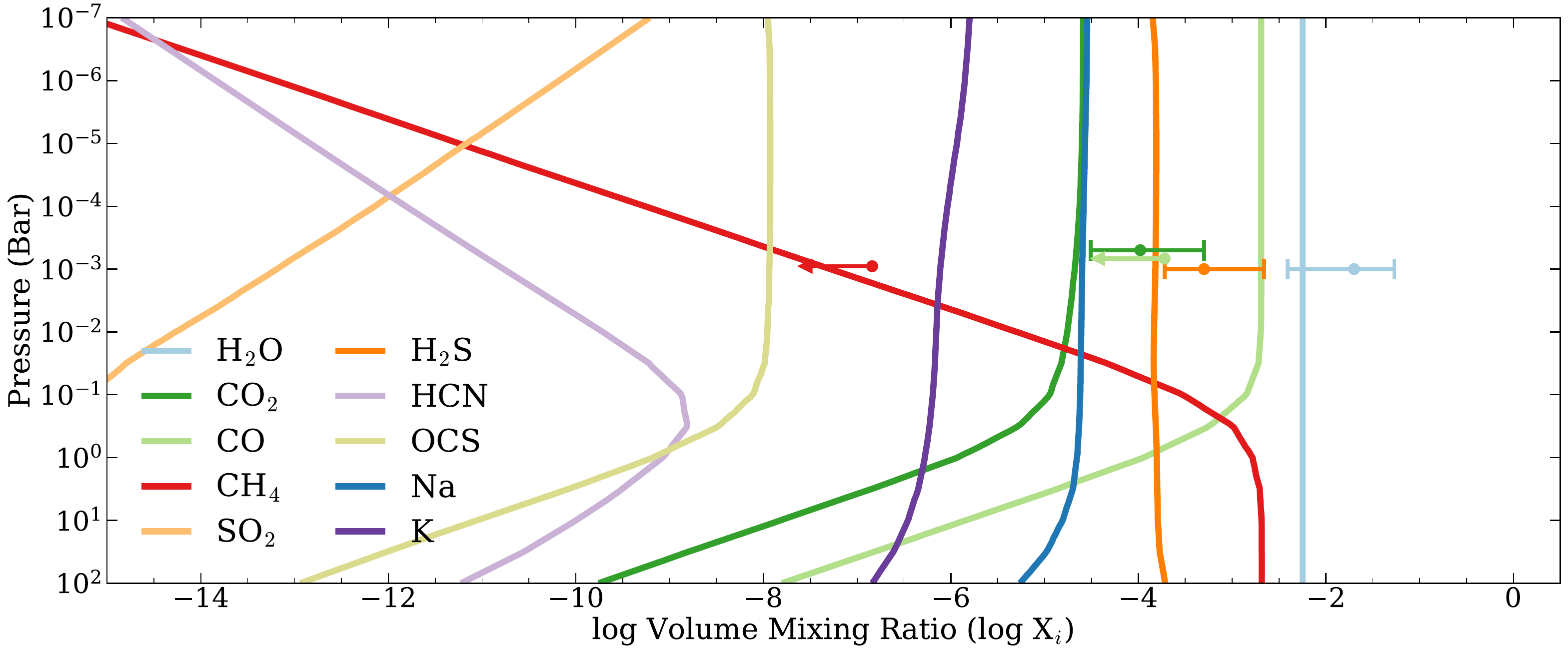}
    \caption{Comparison between the equilibrium chemical profiles and free chemistry volume mixing ratios inferred via our \poseidon retrieval analysis. We plot as the colored lines the equilibrium chemistry volume mixing ratios as a function of pressure, and as colored points the corresponding VMRs from our free retrieval. We plot error bars to represent the 1-$\sigma$ upper and lower bounds for H$_2$O, CO$_2$, and H$_2$S whereas we plot left-pointing arrows to represent the 2-$\sigma$ upper limits for CO and CH$_4$. While our equilibrium chemistry retrieval prefers a low C/O ratio that approaches the lower end of the prior range in conjunction with cloud-muted features, our free retrieval instead prefers a low computed C/O ratio of 0.006 with patchy cloud coverage. This value is calculated from the stoichiometric sum of the freely retrieved molecular VMRs and is not the equilibrium-retrieval C/O posterior shown in \autoref{fig:panchromatic_eq_retrieval}; because CO is not constrained, the free-retrieval value should not be interpreted as a robust elemental abundance ratio. The discrepancy between our equilibrium and free chemistry retrievals indicates that disequilibrium chemistry is affecting \planetname{}'s atmosphere, rendering equilibrium chemical retrievals inadequate for the purpose of estimating the planet's atmospheric properties.}
    \label{fig:chemistry}
\end{figure*}

The discrepancies between our \poseidon equilibrium and free chemistry retrievals could be explained by the presence of disequilibrium chemistry.
As we show in Figure \ref{fig:chemistry}, the freely-retrieved volume mixing ratios correspond to an higher metallicity and a far lower computed C/O than our equilibrium chemistry retrieval (which should not be trusted due to unconstrained CO).
While our inferred VMRs for CO$_2$ and H$_2$S are near the values expected from equilibrium chemistry, H$_2$O and CO are both considerably lower.
While this is in part due to the equilibrium chemistry retrieval preferring more muted features above a terminator-spanning cloud deck, these differences may also indicate that disequilibrium chemistry is playing a role in shaping \planetname{}'s atmosphere.

These results independently support strong detections of both \ce{H$_2$O} and \ce{CO$_2$} in the atmosphere of \planetname{}.
Although the absolute Bayesian evidences should not be compared directly between retrieval frameworks because of differences in model implementation, priors, opacity treatment, and sampling details, the nested model comparisons within each framework lead to the same physical interpretation: the combined HST+JWST spectrum requires \ce{H$_2$O} and \ce{CO$_2$} absorption. 
Thus, the \sansar retrieval is largely in agreement with the \poseidon results, as shown by the comparison of retrieved posteriors from both retrievals in \autoref{tab:vmr_posteriors} and \autoref{fig:overlay_corner}, and the comparison of retrieval statistics in \autoref{tab:stats}.

\section{Discussion}
\label{sec:discuss}

Our panchromatic free chemistry retrieval (\autoref{fig:panchromatic_free_retrieval}), combining the new JWST NIRSpec G395H transmission spectrum with the archival HST spectra, robustly recovers the \ce{H$_2$O} absorption at the terminator of \planetname{}, consistent with the previously reported $>5\sigma$ WFC3 water band detection at $\sim1.4\mu m$ \citep{Wakeford13}. 
Our CO$_2$ detection ($\log_{10}(\mathcal{B})_{{\ce{CO_2}}}=52.3$), driven primarily by absorption in the G395H bandpass, provides significantly tighter constraint relative to predictions made through broadband photometry with Spitzer's IRAC \citep{Todorov2010_Spitzer}. 
The additional wavelength coverage provided by G395H also reduces the reliance or bias associated with a single NIR spectral structure, with our panchromatic multiple-instrument approach lessening the biases of individual dataset-dependent offsets during retrievals. 

Taken together, the retrieved molecular inventory provides a coherent picture of \planetname{} as an oxygen-rich, metal-enriched hot Jupiter whose molecular features are not strongly muted by aerosols. The strong \ce{H2O} and \ce{CO2} constraints, together with the equilibrium chemistry preference for $\log_{10}\mathrm{[M/H]} = 0.99^{+0.19}_{-0.14}$ and a 3$\sigma$ upper limit of C/O $<0.52$, suggest the observable atmosphere is ample in heavy elements while remaining oxygen-dominated. In our retrievals, the 4.3~$\mu$m feature provides the strongest new constraint on the heavy-element enrichment beyond previously known \ce{H2O} absorption. We therefore interpret the inferred metallicity and C/O ratio in the standardized abundance frameworks emphasized by \citet{Fu2025a} and \citet{Lothringer2026}, where consistent definitions of metallicity, C/O, and the adopted abundance scale are essential for comparing planets across the broader exoplanet population. 


 The retrieval comparisons do not provide compelling statistical evidence for either uniform or inhomogeneous clouds: the cloud parameters remain broadly constrained, the Bayes factor is small, and BPICs favors the cloud-free model. We therefore avoid classifying the terminator as cloud free. Instead, the persistence and amplitude of molecular structure across the combined STIS, WFC3, and G395H wavelength range argue specifically against strong aerosol muting. This observational conclusion is consistent with the ``Clear Sky Corridor'' framework of \citet{Ashtari2025}, which predicts relatively unmuted molecular features for planets near \planetname{}'s mass--temperature regime \citep{Wakeford13,Nikolov14}.  

The JWST Exoplanet Grand Tour also provides a uniform framework for studying limb asymmetries in giant exoplanets \citep{Fu2025b, Mukherjee2026}. 
Using NIRISS SOSS observations, \citet{Fu2025b} found several hot Jupiters to show prominent limb-to-limb differences in their transit depth within the 1.4~$\mu$m \ce{H$_2$O} band, with muted morning limbs and clearer evening limbs. 
Within this context, \planetname{}'s equilibrium temperature falls within the demographic of planets with more symmetrical limbs. 
Our observations of \planetname{} support these predictions from \citet{Fu2025b} yielding an unremarkable difference between evening and morning limb spectra (see \autoref{fig:limb_spectra}). 
The modeled trend for limb asymmetries \citep{Fu2025b} is also primarily based on shorter wavelength SOSS observations, whereas the present observations probe the longer wavelength G395H bandpass less sensitive to limb-asymmetries. 
Nevertheless, we performed a limb-resolved analysis for the G395H spectra of \planetname{} for consistency, with our generally null-result for limb asymmetry (\autoref{fig:limb_spectra}) being consistent with predictions from Figure 8 of \citet{Fu2025b} for \planetname{}.

The tentative \ce{H2S} signal should be interpreted in the broader context of sulfur chemistry in giant planet atmospheres.
The presence of \ce{H2S} in a warm or hot giant planet is not, by itself, unexpected: sulfur chemistry models predict \ce{H2S} to be visible over a broad range of hydrogen-dominated atmospheres between the cooler regime where sulfur can be sequestered into condensates and the ultra-hot Jupiter regimes where \ce{H2S} is thermally dissociated \citep{Polman2023, Mukherjee2025}.
Direct observational constraints on \ce{H2S}, however, remain relatively limited. 
Recent JWST observations have begun to change this picture, with \ce{H2S} detected in transmission spectroscopy of HD~189733\,b \citep{Fu2024} and TOI-5205~b \citep{Canas2026}, and inferred from direct-imaging spectroscopy of the HR~8799 planetary system \citep{Ruffio2026}. 

This tentative \ce{H2S} inference also fits into an emerging picture in which the dominant observable sulfur carrier changes across temperature and irradiation regimes \citep{Mukherjee2025}. Recent JWST observations have reported \ce{CS2} in the warm giant WASP-80~b, and in the young, strongly irradiated V1298~Tau~e, while V1298~Tau~b shows \ce{SO2}, highlighting that \ce{H2S}, \ce{SO2}, and \ce{CS2} may trace distinct sulfur-chemistry regimes \citep{Triantafillides2026, Dai2026, Mukherjee2025}. This makes the the weak preference for \ce{H2S} in \planetname{}'s atmosphere qualitatively consistent with sulfur-chemistry predictions for planets in this temperature regime.

Our interpretation of \planetname{}'s atmosphere is consistent with the ``Clear Sky Corridor" framework of \citet{Ashtari2025}. 
In a strongly cloud- or haze-muted atmosphere, the comparatively subtle \ce{H2S} absorption between the broader \ce{H2O} and \ce{CO2} features would be more difficult to detect, and could be absorbed into continuum-level bumps in the spectrum or aerosol degeneracies. 
 The lack of strong aerosol muting may therefore be important for this tentative sulfur inference, because it permits G395H transmission spectroscopy to probe pressures at which \ce{H2S} opacity can contribute measurably. 
If confirmed with future JWST observations, \planetname{} would support the idea that planets with weakly muted molecular spectra are especially favorable targets for sulfur measurements. 
These valuable sulfur abundance ratios provide a complementary compositional axis to metallicity and C/O, and may help probe refractory-to-volatile enrichment pathways in giant-planet formation \citep{Lothringer2021, Crossfield2025}.


We caution, however, that the present \ce{H2S} inference remains tentative. 
Due to the relatively small feature size compared to the pronounced, broader \ce{H2O} and \ce{CO2} opacities present, the \ce{H2S} measurement is more vulnerable to instrument systematics. 
In addition, \ce{H2S} is photochemically coupled to other sulfur-bearing-species, especially \ce{SO2}, and overlapping opacities between these species can complicate retrieved abundance estimates \citep{Tsai2023}. 
Future observations that jointly constrain the 1.4~$\mu$m \ce{H2O} band, the 3-4~$\mu$m sulfur-sensitive region, and the 4.3~$\mu$m \ce{CO2} band will be needed to determine whether \planetname{} truly hosts detectable \ce{H2S} and whether its sulfur abundance can be used as a tracer of refractory-to-volatile enrichment.


\section{Conclusions} 
\label{sec:conclusion}

As part of the JWST Exoplanet Grand Tour program, we present JWST NIRSpec G395H transmission spectroscopy of the hot Jupiter \planetname{}, extending previous optical and near-infrared atmospheric constraints from HST STIS and WFC3 to a panchromatic wavelength range of 0.3--5.3~$\mu$m. 

Atmospheric retrievals of the combined HST$+$JWST transmission spectrum indicate strong evidence for \ce{H$_2$O} and \ce{CO$_2$} in the terminator region of \planetname{}. In free-chemistry retrievals, \ce{H$_2$O} and \ce{CO$_2$} are detected with Bayes factors of $\log_{10}(\mathcal{B})_{{\ce{H_2O}}}=8.9$ and $\log_{10}(\mathcal{B})_{\ce{CO_2}}=52.3$, respectively, while \ce{H$_2$S} is favored at a more tentative level with $\log_{10}(\mathcal{B})_{\ce{H_2S}}=1.4$. The newly added \ce{CO$_2$} constraint is driven primarily by the strong $\sim4.3~\mu$m \ce{CO$_2$} absorption feature made available by this G395H observation, while the \ce{H$_2$O} abundance is informed jointly by the archival WFC3 water band and new NIRSpec coverage. We interpret these abundance constraints relative to the host-star composition, since atmospheric metallicities and C/O ratios are most meaningful when compared to the natal stellar abundance pattern rather than Solar values alone \citep{Reggiani2022}. For this work, we adopt stellar parameters for \planetname{} from the homogeneous stellar analysis of \citet{McCreery2026} for all host stars in the JWST Exoplanet Grand Tour program (GO 5924). Under equilibrium chemistry assumptions, the atmosphere is favored to have  $\log_{10}\mathrm{[M/H]}=0.99^{+0.19}_{-0.14}$, corresponding to approximately $10\times$ Solar or $9\times$ the near-Solar host-star metallicity ($\mathrm{[M/H]}=0.04\pm0.02$), and a subsolar C/O ratio, consistent with an oxygen-rich composition dominated by \ce{H$_2$O} and \ce{CO$_2$} opacity.

 Although the retrievals do not distinguish decisively among cloud prescriptions, the molecular feature amplitudes observed across the HST and JWST wavelength range rule out a strongly cloud-muted terminator. While the present data provides strong constraints on \ce{H$_2$O} and \ce{CO$_2$} and a weak hint of sulfur chemistry via \ce{H$_2$S}, the interpretation of the latter remains subject to limitations of degeneracies in retrievals among clouds, abundances, thermal structure, and overlapping opacities \citep{Tsai2023} (full corner plots shown in \autoref{fig:full_corner_plots}). Future observations, particularly complementary JWST coverage, will help test the tentative sulfur-bearing signal, and further constrain the metallicity, C/O ratio, and cloud properties of \planetname{}. Together, these results establish \planetname{} as a valuable benchmark for comparative studies of hot-Jupiter atmospheric composition.



\begin{acknowledgments}
Stephen P. Schmidt is supported by the National Science Foundation Graduate Research Fellowship Program under Grant No. DGE2139757.
NJM was also supported by a UK Research and Innovation (UKRI) Future Leaders Fellowship MR/T040866/1, and a Science and Technology Facilities Council (STFC) (STFC) Small Award ST/Y00261X/1.

This work is based on observations made with the NASA/ESA/CSA JWST. The data were obtained from the Mikulski Archive for Space Telescopes at the Space Telescope Science Institute, which is operated by the Association of Universities for Research in Astronomy, Inc., under NASA contract NAS 5-03127 for JWST. The specific observations analyzed can be accessed via \dataset[DOI: 10.17909/va0w-hr48]{https://doi.org/10.17909/va0w-hr48}. 

This research made use of the NASA/IPAC Infrared Science Archive, which is funded by the National Aeronautics and Space Administration and operated by the California Institute of Technology.

AI-assisted tools, including ChatGPT by OpenAI, were used for editorial support, refinement of LaTeX formatting and structure, and guidance in debugging \texttt{python} code. These tools were not used for scientific analysis, data interpretation, or the generation of scientific results. All scientific content, methodology, and conclusions remain the sole responsibility of the authors.

Support for JWST program GO-5924 was provided by NASA through a grant from the Space Telescope Science Institute, which is operated by the Association of Universities for Research in Astronomy, Inc., under NASA contract NAS 5-26555. 
This research has made use of the NASA Exoplanet Archive, which is operated by the California Institute of Technology, under contract with the National Aeronautics and Space Administration under the Exoplanet Exploration Program. 
This research has made use of NASA's Astrophysics Data System Bibliographic Services.

\end{acknowledgments}

\vspace{5mm}
\facilities{ADS, Exoplanet Archive, HST(STIS, WFC3), JWST(NIRSpec), MAST}

\software{
\\ \texttt{astropy} \citep{2013A&A...558A..33A,2018AJ....156..123A, AstropyIII}
\\ \texttt{batman} \citep{Kreidberg2015},
\\ \texttt{catwoman} \citep{Jones22, Espinoza21},
\\ \texttt{emcee} \citep{Foremak-Mackey2013},
\\ \eureka \citep{Bell2022},
\\ \texttt{ExoTiC-LD} \citep{grant2024exoticLDJoss},
\\ \firefly \citep{Rus22, Rus23},
\\ IPython \citep{ipython},
\\ \texttt{lacosmic} \citep{lacosmic},
\\ \texttt{lmfit} \citep{lmfit},
\\ \texttt{matplotlib} \citep{hunter2007matplotlib},
\\ \texttt{numpy} \citep{harris2020array},
\\ \poseidon \citep{MacDonald2017, MacDonald2023},
\\ \texttt{PyMultiNest} \citep{Feroz2009,Buchner2014},
\\ \texttt{scipy} \citep{2020SciPy-NMeth}
          }

\newpage
\clearpage
\appendix
\setcounter{figure}{0}
\renewcommand{\thefigure}{A\arabic{figure}}\renewcommand{\theHfigure}{A\arabic{figure}}
\setcounter{table}{0}
\renewcommand{\thetable}{A\arabic{table}}\renewcommand{\theHtable}{A\arabic{table}}
\onecolumngrid

\section{Additional Data Analysis Materials} 
\label{appendix:more_data_analysis}

\begin{figure}[b!]
    \centering
    \includegraphics[width=1.0\linewidth, trim={0cm 0cm 0cm 0cm},clip]{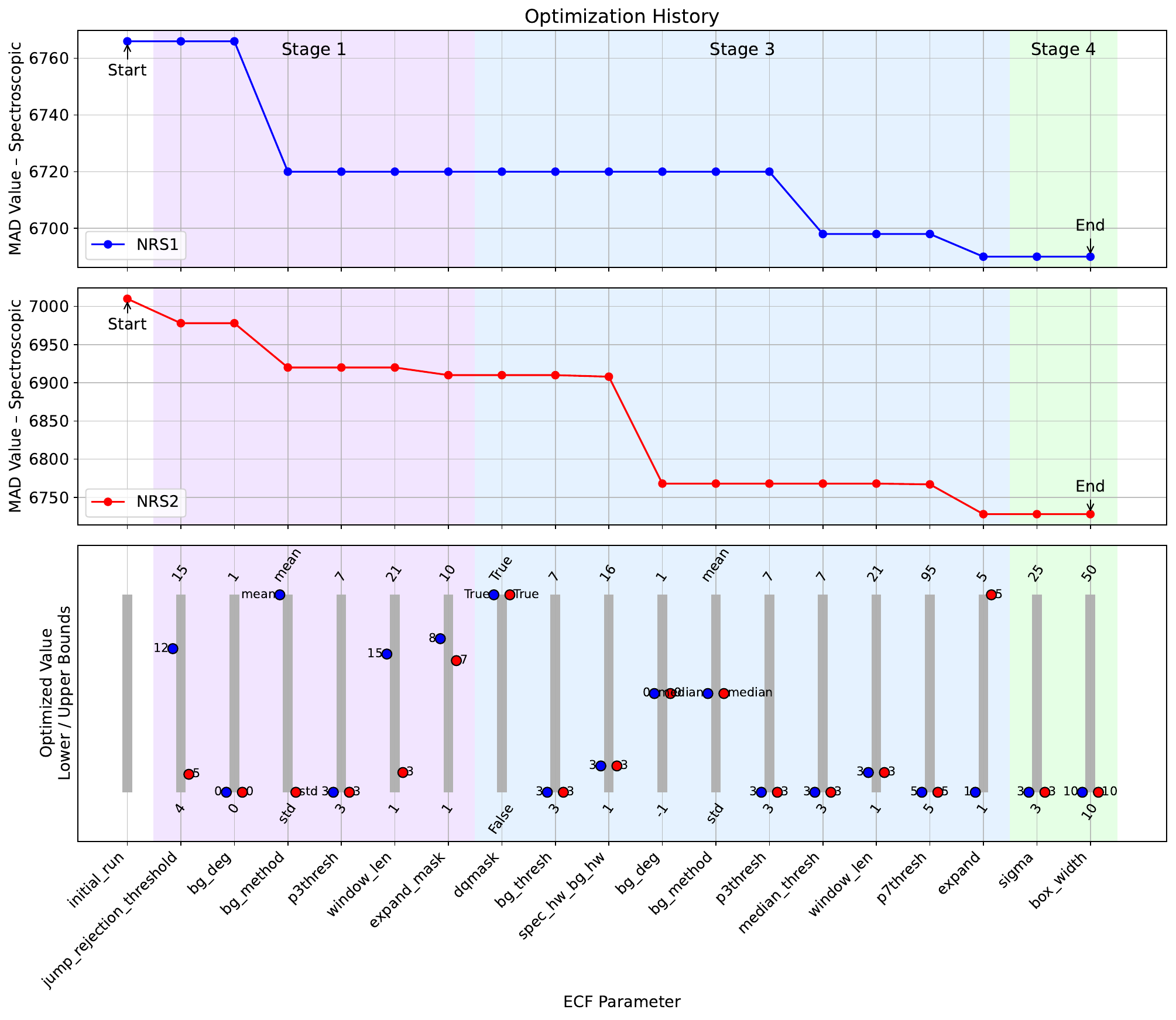} 
    \caption{Optimization history for the parameter-by-parameter data reduction performed with the \eureka pipeline. 
\emph{Top and middle panels}: evolution of the median absolute deviation (MAD) of the spectroscopic light curves for the NRS1 (blue) and NRS2 (red) detectors as successive reduction parameters are optimized. The MAD values are reported in ppm. Parameters are varied sequentially, and the value that minimizes the MAD metric is adopted before proceeding to the next step. Shaded regions indicate groups of related optimization stages. The overall decrease in MAD demonstrates progressive improvement in light-curve quality without evidence for instability or over-tuning. 
\emph{Bottom panel}: optimized parameter values selected at each step, shown relative to their allowed bounds (gray bars), with blue and red markers denoting the final adopted values for NRS1 and NRS2, respectively. Together, these panels illustrate the process and convergence of the optimization framework used to derive the final \eureka transmission spectrum.
}
    \label{fig:optimization_history}
\end{figure}

\begin{deluxetable*}{llrlrlrlrl}[htb!]

\tabletypesize{\scriptsize}
\tablewidth{0pt}
\tablecaption{Best-fit orbital and transit parameters derived for the three independent JWST reductions (\eureka, \firefly, and \Tswift), together with the previous joint HST/STIS+WFC3 system parameters from \citet{Nikolov14}. All reductions yield mutually consistent system parameters within uncertainties.\label{tab:orbital_parameters}}
\tablehead{
    Parameter & Unit &
    \multicolumn{2}{c}{\eureka} & 
    \multicolumn{2}{c}{\firefly} & 
    \multicolumn{2}{c}{\Tswift} &
    \multicolumn{2}{c}{HST}
}
\startdata
R$_p$/R$_s$ NRS1 & \nodata & 0.117651 & $\pm$3.9$\cdot10^{-5}$ & 0.118064 & $\pm$1.7$\cdot10^{-4}$ & 0.118031 & $\pm$1.8$\cdot10^{-4}$ & 0.11802 & $\pm$1.8$\cdot10^{-4}$ \\
R$_p$/R$_s$ NRS2 & \nodata & 0.117682 & $\pm$3.9$\cdot10^{-5}$ & 0.117936 & $\pm$2.5$\cdot10^{-4}$ & 0.117748 & $\pm$2.0$\cdot10^{-4}$ & \nodata & \nodata \\
Period, $P$ & d & 4.4652934 & (fixed) & 4.4653 & (fixed) & 4.4652998 & (fixed) & 4.46529976 & $\pm5.5\cdot10^{-7}$ \\
Transit time $T_0-2460000$ & BJD$_{\rm TDB}$ d & 633.22878 & $\pm2\cdot10^{-5}$ & 633.228843 & $\pm1.6\cdot10^{-5}$ & 633.228864 & $\pm1.2875\cdot10^{-5}$ & \nodata & \nodata \\
$a/R_\star$ & \nodata & 9.975 & $\pm$0.019 & 9.961 & $\pm$0.017 & 9.972988 & $\pm$0.014 & 9.853 & $\pm$0.071 \\
impact parameter, $b$ & \nodata & \nodata & \nodata & 0.7445 & $\pm$0.0012 & \nodata & \nodata & 0.7501 & $^{+0.0064}_{-0.0069}$ \\
inclination, $i$ & \textdegree & 85.728 & $\pm$0.015 & \nodata & \nodata & 85.725 & $\pm$0.01177 & 85.634 & $\pm$0.056 \\
limb darkening, $u_1$ NRS1 & \nodata & 0.130 & $\pm$0.034 & 0.148 & $\pm$0.038 & 0.139496 & $\pm$0.045537 & \nodata & \nodata \\
limb darkening, $u_2$ NRS1 & \nodata & 0.063 & $\pm$0.035 & 0.030 & $\pm$0.038 & 0.050385 & $\pm$0.034964 & \nodata & \nodata \\
limb darkening, $u_1$ NRS2 & \nodata & 0.143 & $\pm$0.042 & 0.102 & $\pm$0.06 & 0.139228 & $\pm$0.053297 & \nodata & \nodata \\
limb darkening, $u_2$ NRS2 & \nodata & 0.073 & $\pm$0.042 & 0.079 & $\pm$0.06 & 0.050332 & $\pm$0.034734 & \nodata & \nodata \\
\hline
\enddata
\tablecomments{The limb-darkening parameters and uncertainties reported for \firefly have been transformed from the $u_+=u_1+u_2$, $u_-=u_1-u_2$ parameterization. The \firefly radius ratios have likewise been transformed from the squared radius ratio fitted by that pipeline. The smaller formal white-light $R_p/R_\star$ uncertainties from \eureka principally reflect differences among the pipeline likelihoods, uncertainty rescaling, parameter covariances, and the optimized reduction; the present comparison does not isolate the contribution of any one choice. The more similar spectroscopic uncertainties in \autoref{fig:comparison_spectra} are set by the scatter and photon noise within each wavelength channel, so the white-light uncertainty ratio should not be interpreted as a direct ranking of the spectroscopic precision. The HST reference epoch is omitted because it is not reported on the same epoch convention used here.}
\end{deluxetable*}

\begin{table*}[t!]
\centering
\caption{Definitions of the \eureka parameters varied in the sequential optimization shown in \autoref{fig:optimization_history}.\label{tab:eureka_optimizer_parameters}}
{
\footnotesize
\renewcommand{\arraystretch}{1.18}
\begin{tabular}{ll}
\hline
Parameter & Definition \\
\hline
\texttt{jump\_rejection\_threshold} & \parbox[t]{0.58\linewidth}{Sigma threshold used by the Stage~1 jump-detection step.} \\
\texttt{bg\_deg} & \parbox[t]{0.58\linewidth}{Polynomial order of the column-by-column background model.} \\
\texttt{bg\_method} & \parbox[t]{0.58\linewidth}{Scatter estimator used to identify background outliers (standard deviation, median absolute deviation, or mean absolute deviation).} \\
\texttt{p3thresh} & \parbox[t]{0.58\linewidth}{Sigma threshold for rejecting outliers during background subtraction.} \\
\texttt{window\_len} & \parbox[t]{0.58\linewidth}{Smoothing-window length used to construct the median spatial profile; it does not smooth the final spectrum.} \\
\texttt{expand\_mask} & \parbox[t]{0.58\linewidth}{Number of pixels by which the Stage~1 trace mask is expanded.} \\
\texttt{dqmask} & \parbox[t]{0.58\linewidth}{Whether pixels carrying ``do not use'' data-quality flags are masked.} \\
\texttt{bg\_thresh} & \parbox[t]{0.58\linewidth}{Temporal sigma-clipping threshold applied to background pixels.} \\
\texttt{spec\_hw}, \texttt{bg\_hw} & \parbox[t]{0.58\linewidth}{Half-widths of the extraction aperture and background-exclusion region, respectively, in pixels.} \\
\texttt{median\_thresh} & \parbox[t]{0.58\linewidth}{Sigma threshold for rejecting outliers from the median frame used to construct the spatial profile.} \\
\texttt{p7thresh} & \parbox[t]{0.58\linewidth}{Sigma threshold for rejecting outliers during optimal spectral extraction.} \\
\texttt{expand} & \parbox[t]{0.58\linewidth}{Cross-dispersion supersampling factor.} \\
\texttt{sigma} & \parbox[t]{0.58\linewidth}{Sigma threshold for Stage~4 rolling-median time-series clipping.} \\
\texttt{box\_width} & \parbox[t]{0.58\linewidth}{Width of the Stage~4 rolling-median filter, in integrations.} \\
\hline
\multicolumn{2}{l}{\parbox{0.92\textwidth}{\emph{Note.} The optimized \texttt{bg\_method} choices for NRS1 and NRS2 select different estimators of the background-pixel scatter for outlier rejection; they do not impose different background levels. We therefore do not expect this choice alone to introduce a relative detector offset, consistent with the continuous spectrum and agreement of the detector white-light depths.}} \\
\hline
\end{tabular}
}
\end{table*}

\begin{figure*}[htb!]
    \centering
    \includegraphics[width=0.99\linewidth, trim={0cm 0cm 0cm 0cm},clip]{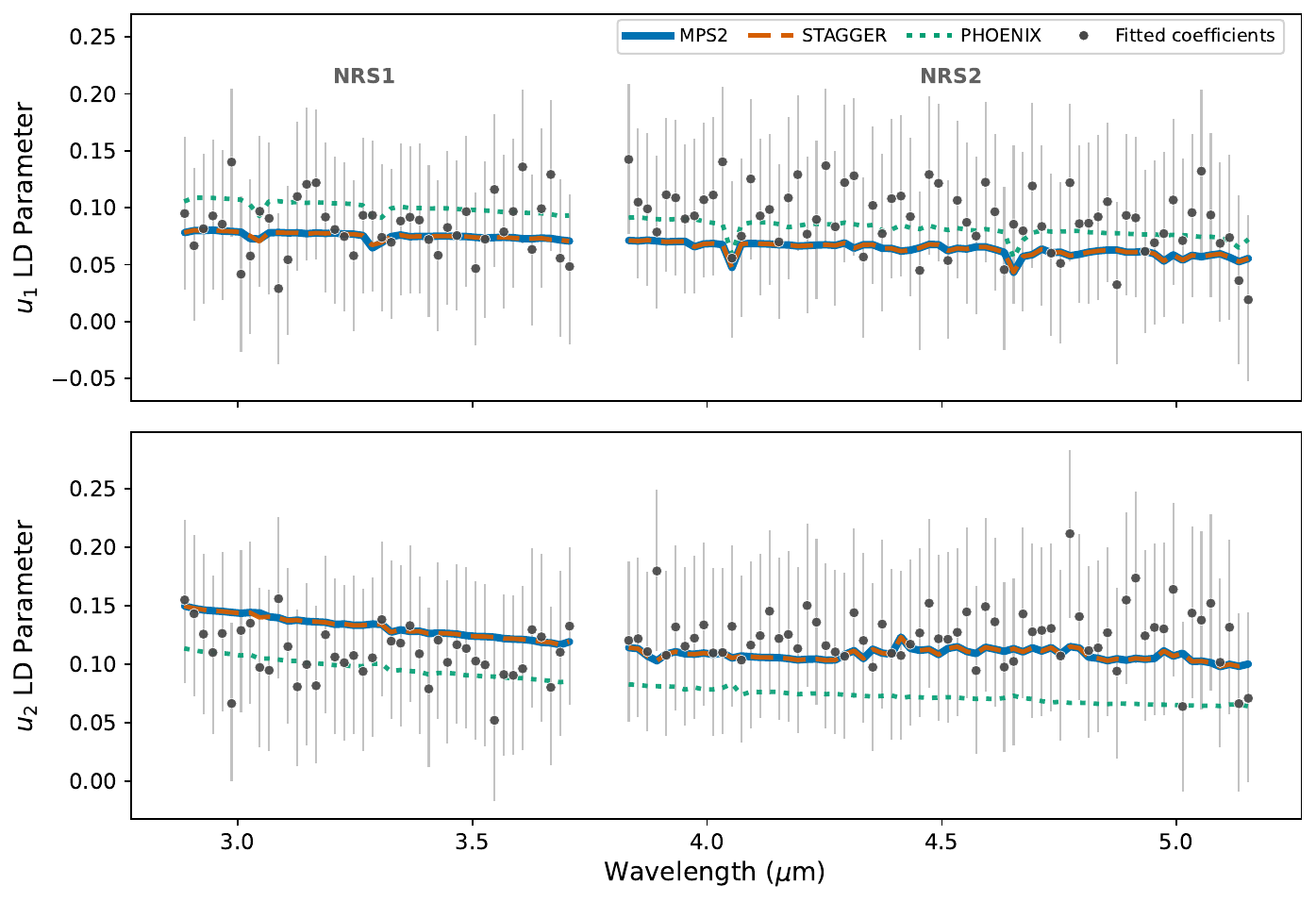} 
    \caption{
    Comparison of wavelength-dependent quadratic limb-darkening coefficients
    ($u_1$, top; $u_2$, bottom)
    predicted by the ExoTiC-LD MPS2, STAGGER, and PHOENIX stellar-atmosphere grids and obtained from model-independent fits to the \eureka{} reduction
    for the NRS1 and NRS2 channels.
    Solid blue, dashed orange, and dotted green curves show the MPS2, STAGGER, and PHOENIX predictions, respectively, while gray circles with error bars show the independently fitted limb-darkening coefficients. For the revised analysis, $u_1$ and $u_2$ are fitted freely in each wavelength channel using uniform priors, $\mathcal{U}(0,1)$, with both coefficients initialized at $u_1=u_2=0.2$. None of the stellar-atmosphere grids is used to initialize, center, or constrain the fitted coefficients; the model curves are shown only as post hoc comparisons to the same freely fitted values.
    Across the full $\sim2.7$--$5.3\,\mu\mathrm{m}$ wavelength range, MPS2 yields reduced $\chi_\nu^2$ values of 1.44 and 1.08 for $u_1$ and $u_2$, respectively, corresponding to a mean value of 1.26. STAGGER provides nearly identical agreement, with values of 1.45 and 1.09 and a mean of 1.27, while PHOENIX yields a mean $\chi_\nu^2$ of 1.86. MPS2 is therefore preferred as it provides the lowest mean $\chi_\nu^2$, although only marginally lower than STAGGER. Because the coefficients used in the light-curve analysis are fitted freely, this comparison identifies the grid that most closely reproduces the empirical coefficients but does not bias a particular stellar-atmosphere grid.
    }

    \label{fig:ld_comparison}
\end{figure*}

\begin{figure*}[htb!]
    \centering
    \includegraphics[width=0.95\linewidth, trim={0cm 0cm 0cm 0cm}, clip]{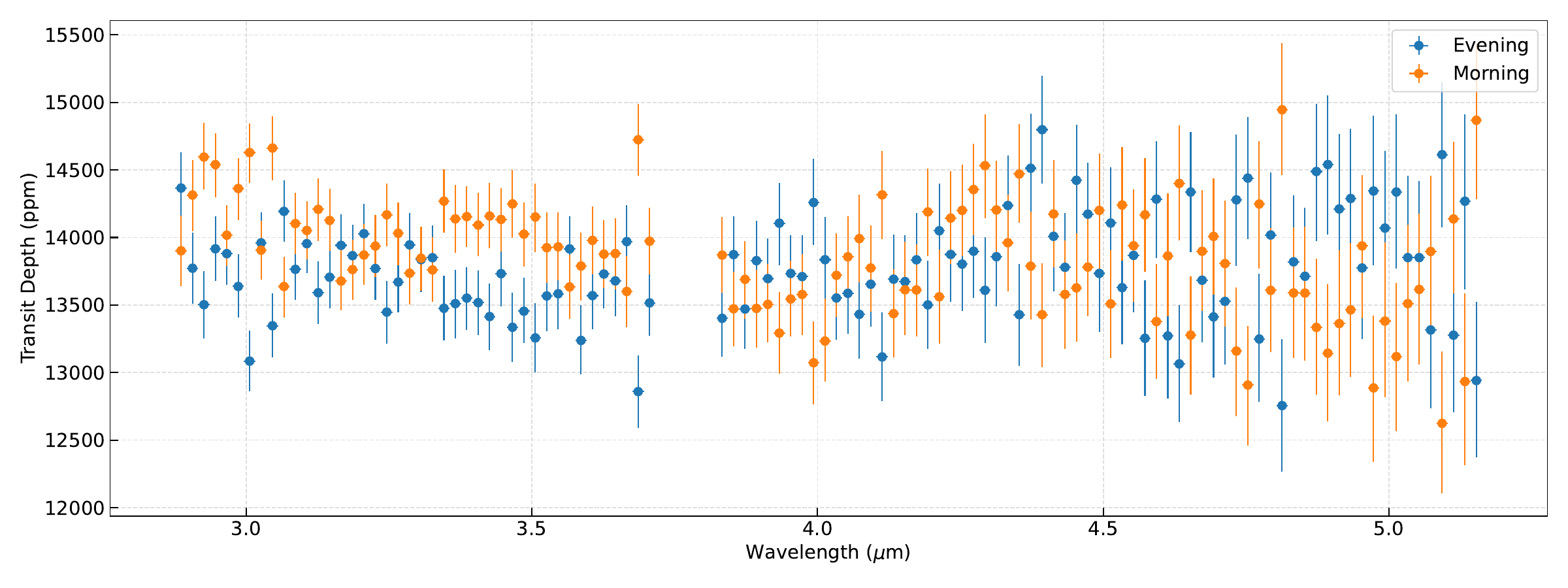}
    \caption{
    Morning- and evening-limb transmission spectra of \planetname{} derived from the JWST NIRSpec G395H observation. Blue points show the evening-limb spectrum and orange points show the morning-limb spectrum, with 1$\sigma$ uncertainties. The two limb spectra are broadly consistent across the G395H bandpass, with no clear wavelength-dependent offset or coherent detector-wide difference. This lack of a significant G395H limb contrast is consistent with expectations from \citet{Fu2025b}, which predicts only a modest morning-evening difference for \planetname{} and identifies the NIRISS SOSS 1.4~$\mu$m \ce{H$_2$O} band as the more sensitive cue for the population-level limb-asymmetry trend.
    }
    \label{fig:limb_spectra}
\end{figure*}

\FloatBarrier
\section{Additional Retrieval Materials} 
\label{appendix:more_retrieval}
\setcounter{figure}{0}
\renewcommand{\thefigure}{B\arabic{figure}}\renewcommand{\theHfigure}{B\arabic{figure}}
\setcounter{table}{0}
\renewcommand{\thetable}{B\arabic{table}}\renewcommand{\theHtable}{B\arabic{table}}

\begin{deluxetable*}{lccc}
\centering
\tablewidth{0pt}
\tablecaption{Atmospheric Retrieval Configuration Comparison. System properties are adopted from the homogeneous Grand Tour analysis of \citet{McCreery2026}.}
\tablehead{
Model Setting & \colhead{\texttt{POSEIDON}} & \colhead{\texttt{SANSAR}}
}
\startdata
    \multicolumn{3}{c}{\textbf{System Properties}} \\
    \hline
    Stellar Radius ($\text{R}_{\mathrm{\ast}}/\text{R}_{\odot}$)  & 1.22 & 1.22 \\
    Planetary Radius ($\text{R}_{\mathrm{p}}/\text{R}_{\text{Jup}}$) & 1.397 & 1.397 \\
    Planetary Mass ($\text{M}_{\mathrm{p}}/\text{M}_{\text{Jup}}$) & 0.525 & --- \\
    Planetary Surface Gravity ($\mathrm{m\,s^{-2}}$) & 6.66 & 6.667\\
    \hline
    \multicolumn{3}{c}{\textbf{Atmospheric Model}}\\
    \hline
    Pressure Grid & $10^{-7}$--$10^2$\,bar & $10^{-8}$--$10^2$\,bar \\
    Number of Layers & 150 & 100 \\
    Molecules Included & H$_2$O, CO$_2$, CH$_4$, CO, HCN,  & H$_2$O, CO$_2$, CH$_4$, CO, HCN, \\
    & SO$_2$, H$_2$S, OCS, Na, K & SO$_2$, H$_2$S, OCS, Na, K \\
    He/H$_2$ Ratio & 0.17 & 0.178 \\
    P-T Profile Treatment & \citet{Madhusudhan2009} & \citet{Madhusudhan2009}\\
    Cloud Treatment & Patchy Cloud Deck \& Haze & Patchy Cloud Deck \& Haze \\
    \hspace{1pt} (citation) & \citet{MacDonald2017} & ---\\
    Hydrostatic Boundary Condition & $\log_{\rm{10}} P_{\rm{ref}}$ & $\log_{\rm{10}} P_{\rm{ref}}$ \\
    \hline
    \multicolumn{3}{c}{\textbf{Spectral Model}}\\
    \hline
    Model Wavelength Grid & 0.3--5.3\,$\mu\text{m}$ & 0.3--5.2\,$\mu\text{m}$ \\
    Native Opacity Resolution & 0.01\,cm$^{-1}$ & 0.001\,cm$^{-1}$ \\
    Opacity Sampling Resolution & $R = $ 20,000 & --- \\
    Correlated-k Resolution & --- & $R = $ 1,000\\
    \hline
    \multicolumn{3}{c}{\textbf{Retrieval Priors and Settings}}\\
    \hline
    Planetary Mass (M$_{\rm{p}}/\text{M}_{\text{Jup}}$) & 0.525  (Fixed) & ---\\
    Planetary Surface Gravity ($\log_{10}[g_{\rm p}/\mathrm{cm\,s^{-2}}]$) & 2.82 (Fixed) & 2.824 (Fixed) \\
    Reference Pressure ($\log_{\rm{10}} (P_{\rm{ref}}$ / bar)) & 10$^{-3}$ (fixed) & 10$^{-2}$ (fixed) \\
    Reference Radius ($\text{R}_{\mathrm{p, \, ref}}/\text{R}_{\text{Jup}}$) & $\mathcal{N}(1.397, 0.279^2)$ & $\mathcal{U}(1.0, 1.6)$ \\
    \hline
    Atmospheric Temperature ($T_{\mathrm{ref}}$) & $\mathcal{U}(750, 2500)$\,K & $\mathcal{U}(600, 2000)$\,K \\
    P-T Profile Curvature 1 ($\alpha_{1}$) & $\mathcal{U}(0.3, 2.00)$\,K$^{-\frac{1}{2}}$ & $\mathcal{U}(0.02, 2.00)$\,K$^{-\frac{1}{2}}$ \\
    P-T Profile Curvature 2 ($\alpha_{2}$) & $\mathcal{U}(0.3, 2.00)$\,K$^{-\frac{1}{2}}$ &  $\mathcal{U}(0.02, 2.00)$\,K$^{-\frac{1}{2}}$  \\
    P-T Profile Region 1 ($\log_{10} (P_{1}$ / bar)) & $\mathcal{U}(-6, 0)$ &  $\mathcal{U}(-8, 2)$ \\
    P-T Profile Region 2 ($\log_{10} (P_{2}$ / bar)) & $\mathcal{U}(-6, 0)$ &  $\mathcal{U}(-8, 2)$ \\
    P-T Profile Region 3 ($\log_{10} (P_{3}$ / bar)) & $\mathcal{U}(-2, 2)$ &  $\mathcal{U}(-4, 2)$  \\
    Species VMR ($\log_{\rm{10}} X_i$) & $\mathcal{U}(-12, -0.3)$ & $\mathcal{U}(-12, -1.0)$ \\
    Metallicity ($\log_{\rm{10}} \text{[Z/H]}$) & $\mathcal{U}(0.5, 4.0)$ & --- \\
    C/O & $\mathcal{U}(0.2, 2.0)$ & --- \\
    Cloud Top Pressure ($\log_{\rm{10}} (P_{\rm{cloud}}$ / bar)) & $\mathcal{U}(-5, 0.5)$ & $\mathcal{U}(-8, 2)$ \\
    Cloud Fraction ($\bar{\phi}$) & $\mathcal{U}(0, 1)$ & $\mathcal{U}(0, 1)$ \\
    Haze Rayleigh Enhancement ($\log_{10} a$) & $\mathcal{U}(-4, 8)$ &  $\mathcal{U}(-4, 12)$ \\
    Haze Slope ($\gamma$) & $\mathcal{U}(-20, 2)$ & $\mathcal{U}(-20, -2)$ \\
    Data Offset 1 ($\delta_{\mathrm{rel, \, 1}}$)(ppm) & ---& $\mathcal{U}(-500, 500)$  \\
    Data Offset 2 ($\delta_{\mathrm{rel, \, 2}}$)(ppm) & --- & $\mathcal{U}(-500, 500)$  \\
    \texttt{MultiNest} Live Points & 1000 & 1000  \\
    Retrieval Code Availability & \href{https://github.com/MartianColonist/POSEIDON}{\texttt{POSEIDON} GitHub} & --- \\
    \hline
\enddata
\tablecomments{Gaussian priors are summarized as $\mathcal{N}(\mu, \sigma^2)$, where $\mu$ and $\sigma$ are the mean and standard deviation, respectively. $T_{\mathrm{ref}}$ refers to the top-of-atmosphere temperature for \texttt{POSEIDON}. The stellar radius, planetary radius, planetary mass, and surface gravity are taken from \citet{McCreery2026}.}
\label{tab:retrieval_configurations}
\end{deluxetable*}

\begin{figure*}[htb!]
    \centering
    \includegraphics[width=1.0\linewidth, trim={3.0cm 0.4cm 2.3cm 0.2cm},clip]{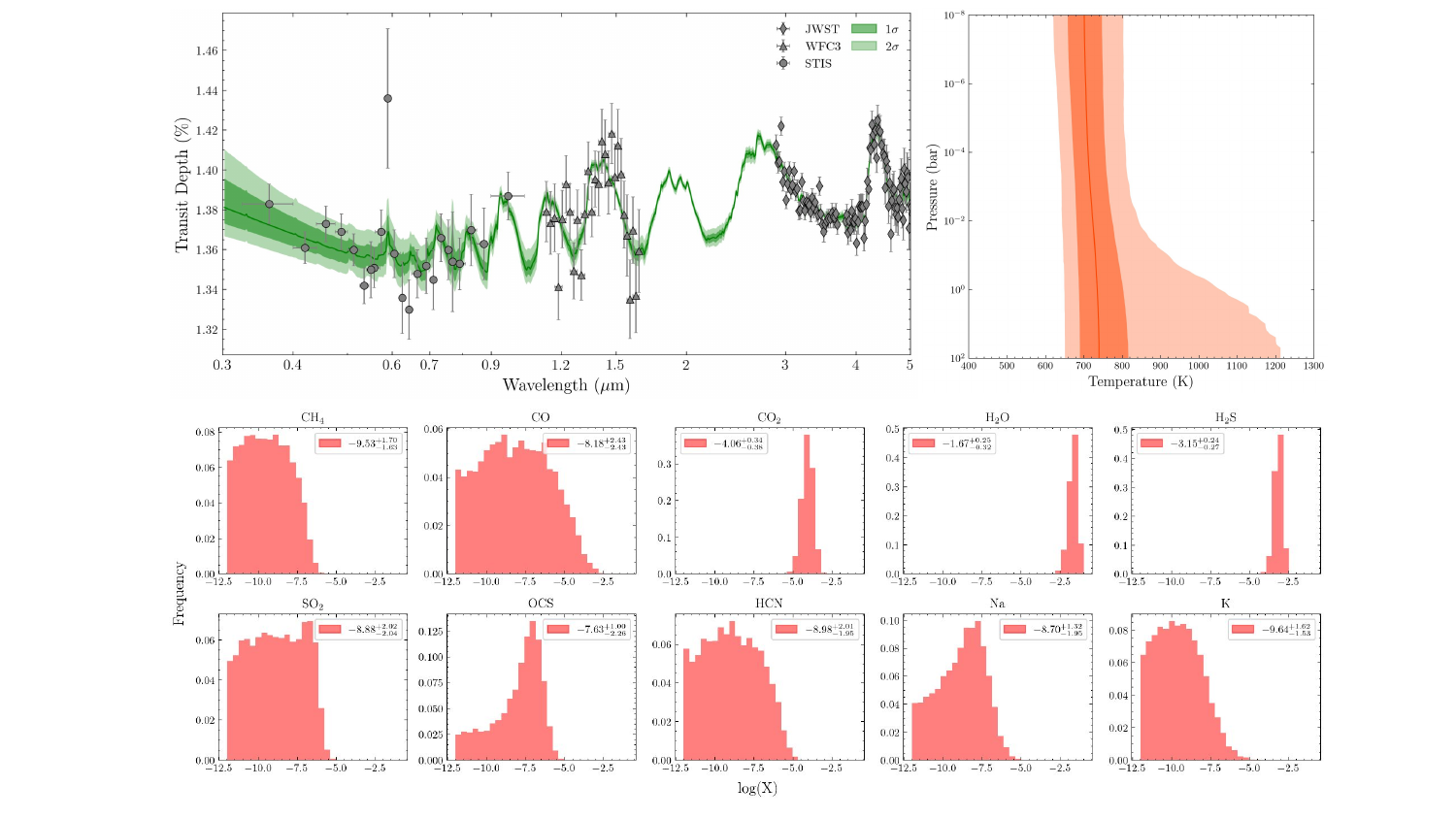}
    \caption{Summary of the independent \sansar free-chemistry retrieval. The upper-left panel compares the median retrieved transmission spectrum and its 1$\sigma$ and 2$\sigma$ credible intervals with the HST/STIS, HST/WFC3, and JWST/NIRSpec observations. The upper-right panel shows the retrieved pressure--temperature profile and credible intervals. The lower panels show the marginalized posterior distributions for the molecular volume mixing ratios; narrow posteriors are recovered for \ce{H2O}, \ce{CO2}, and \ce{H2S}, while the remaining species are weakly constrained.}
    \label{fig:sansar_panchromatic_retrieval}
\end{figure*}

\begin{figure*}[htb!]
    \centering
    \includegraphics[width=1.0\linewidth, trim={0cm 0cm 0cm 0cm},clip]{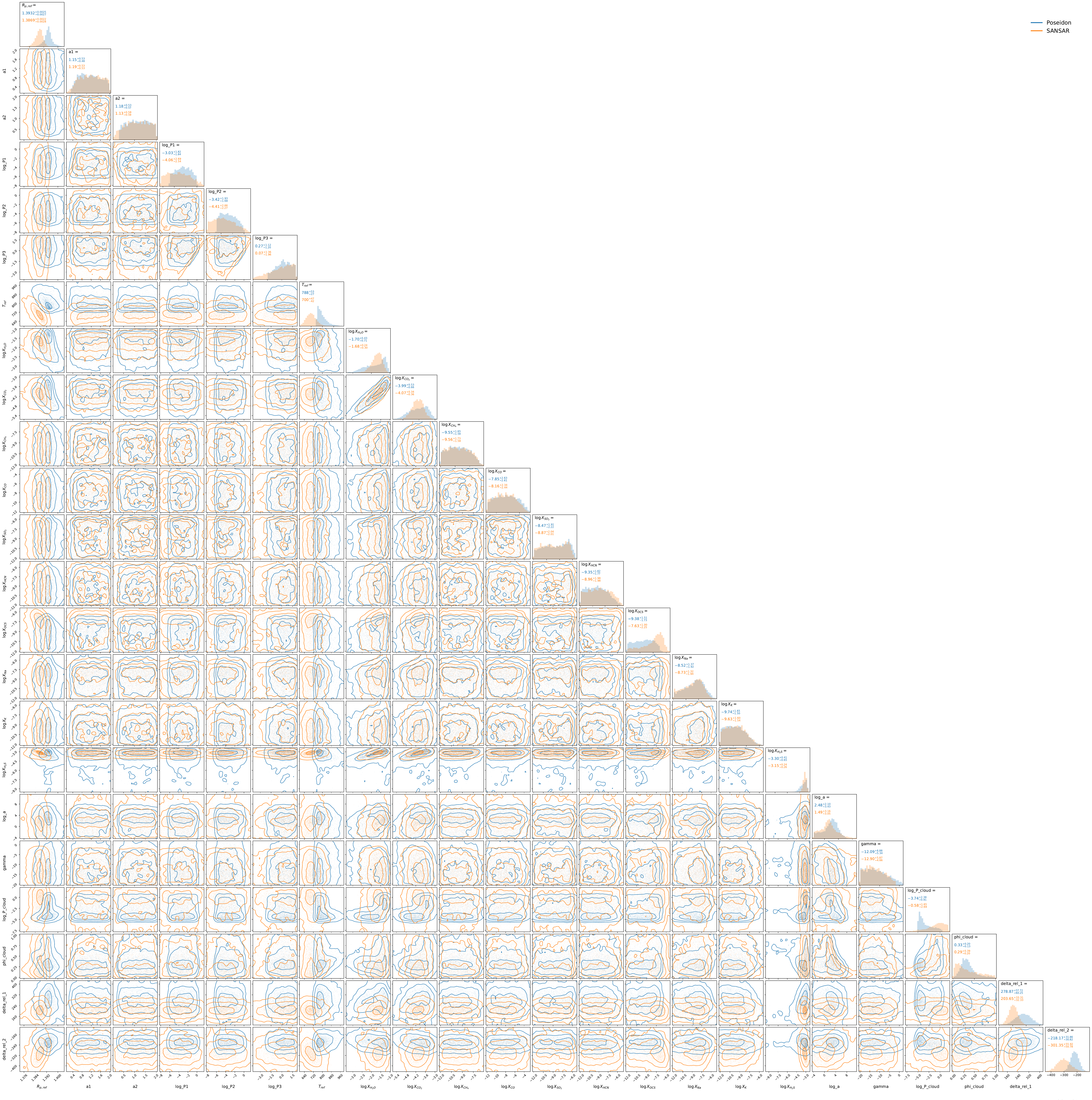}
    \caption{Overlaid corner plot for the full posterior distributions of both \poseidon and \sansar retrievals.}
    \label{fig:full_corner_plots}
\end{figure*}

\FloatBarrier
\bibliography{HATP1b_G395H_rev1}{}
\bibliographystyle{aasjournal}

\end{document}